\documentclass{svjour3}                     
\smartqed  
\usepackage{graphicx}
\usepackage{subfigure}
\usepackage[rightcaption]{sidecap}
\sidecaptionvpos{figure}{t}
\usepackage{threeparttable}
\usepackage{natbib}
\usepackage{xspace}
\usepackage{amsmath}	
\usepackage{amssymb}	

\newcommand{\alfven}{Alfv\'en\xspace}
\newcommand{\unit}[1]{\ensuremath{\,\mathrm {#1}}}  
\renewcommand{\d}{\ensuremath{\mathrm{d}}}  
\newcommand{\A}{\ensuremath{\mathrm{A}}}  %
\newcommand{\s}{\ensuremath{\mathrm{s}}} 
\newcommand{\T}{\ensuremath{\mathrm{T}}}  
 
\renewcommand{\i}{\ensuremath{\mathrm{i}}}  
\newcommand{\e}{\ensuremath{\mathrm{e}}} 
\newcommand{\figref}[1]{Figure~\ref{#1}}

\renewcommand{\eqref}[1]{Equation~(\ref{#1})}
\newcommand{\bvec}[1]{\ensuremath{\boldsymbol{#1}}}  
\renewcommand{\div}[1]{\nabla\cdot #1} 
\newcommand{\curl}[1]{\nabla\times #1} 
\newcommand{\grad}[1]{\nabla#1} 
\newcommand{\RR}{\ensuremath{\mathcal{R}}}
\newcommand{\singlequote}[1]{\lq{#1}\rq}

\renewcommand{\deg}{\ensuremath{^\circ}}

\begin{document}

\title{Slow-Mode Magnetoacoustic Waves in Coronal Loops
}

\titlerunning{Slow Mode Waves in Coronal Loops}        

\author{Tongjiang Wang         \and
        Leon Ofman             \and   
        Ding Yuan		 \and
        Fabio Reale		 \and
        Dmitrii Y. Kolotkov	 \and
        Abhishek K. Srivastava 
}


\institute{T. Wang \at
              The Catholic University of America and NASA Goddard Space Flight Center, Code 671, Greenbelt, MD 20771, USA \\
              \email{tongjiang.wang@nasa.gov}           
           \and
           L. Ofman \at
              The Catholic University of America and NASA Goddard Space Flight Center, Code 671, Greenbelt, MD 20771, USA \\
              \email{ofman@cua.edu} \\
      	\and
          D. Yuan \at 
       	   Institute of Space Science and Applied Technology,
	   Harbin Institute of Technology, Shenzhen,
	   Guangdong 518055, China \\
            \email{yuanding@hit.edu.cn} \\
	\and
	F. Reale \at
	   Dipartimento di Fisica \& Chimica, Universit\'{a} di Palermo, Piazza del Parlamento, I-90134 Palermo, Italy \\
  	   \email{fabio.reale@unipa.it} \\
	\and
	D. Y. Kolotkov \at
	   Centre for Fusion, Space and Astrophysics, Physics Department, University of Warwick, Coventry CV4 7AL, United Kingdom \\
           Institute of Solar-Terrestrial Physics SB RAS, Irkutsk 664033, Russia\\
            \email{D.Kolotkov.1@warwick.ac.uk} \\
	\and
	A. K. Srivastava \at
	   Department of Physics, Indian Institute of Technology (BHU), Varanasi-221005, India \\
	 \email{asrivastava.app@itbhu.ac.in}
}

\date{Received: 2020 October 12 / Accepted: 2021 February 19}

\maketitle

\begin{abstract}
Rapidly decaying long-period oscillations often occur in hot coronal loops of active regions associated with small (or micro-) flares. This kind of wave activity was first discovered with the SOHO/SUMER spectrometer from Doppler velocity measurements of hot emission lines, thus also often called ``SUMER" oscillations. They were mainly interpreted as global (or fundamental mode) standing slow magnetoacoustic waves. In addition, increasing evidence has suggested that the decaying harmonic type of pulsations detected in light curves of solar and stellar flares are likely caused by standing slow-mode waves. The study of slow magnetoacoustic waves in coronal loops has become a topic of particular interest in connection with coronal seismology. We review recent results from SDO/AIA and Hinode/XRT observations that have detected both standing and reflected intensity oscillations in hot flaring loops showing the physical properties (e.g., oscillation periods, decay times, and triggers) in accord with the SUMER oscillations. We also review recent advances in theory and numerical modeling of slow-mode waves focusing on the wave excitation and damping mechanisms. MHD simulations in 1D, 2D and 3D have been dedicated to understanding the physical conditions for the generation of a reflected propagating or a standing wave by impulsive heating. Various damping mechanisms and their analysis methods are summarized. Calculations based on linear theory suggest that the non-ideal MHD effects such as thermal conduction, compressive viscosity, and optically thin radiation may dominate in damping of slow-mode waves in coronal loops of different physical conditions. Finally, an overview is given of several important seismological applications such as determination of transport coefficients and heating function. 

\keywords{Solar activity \and Solar corona \and Coronal loops \and Oscillations and waves}
\end{abstract}

\section{Introduction}
\label{intro}
Recent solar observations by high-resolution imaging space telescopes and spectrometers have confirmed that magnetic structures of the solar corona can support a wide range of magnetohydrodynamic (MHD) waves \citep[see recent reviews by][]{liuw14,wan16}. These waves are natural carriers of energy and so may be an important source for coronal heating \citep[see e.g., the reviews by][this issue]{dem15, vand20}. These waves are also important for their relation to the local plasma parameters of the medium allowing a {\it coronal seismology} \citep[e.g.][]{nak2005,dem2012,nak2016}. In particular, the slow magnetoacoustic mode as one of the main types of MHD wave modes present in coronal loops has become the focus of attention. There is considerable observational evidence for the occurrence of slow (propagating and standing) MHD waves in the solar coronal structures. 

Persistently propagating intensity disturbances, first detected in coronal plumes \citep{ofm97,def98} and then in coronal loops \citep{berg99,dem2006}, were originally identified as slow magnetoacoustic waves \citep{ofm99,nak00}. However, recent observations from Hinode/EIS and IRIS revealed that these disturbances are closely associated with intermittent outflows and spicules produced at loop footpoint regions leading to their interpretations in debate (e.g. \citealt{dep2010,wan2012a,wan2012b}, and a detailed discussion in \citealt{wan16}). Flare-induced Doppler velocity oscillations, first discovered in hot active region (AR) loops by the SOHO/SUMER spectrometer (often called ``SUMER" oscillations; \citealt{wan02,wan03a,wan03b}), were interpreted as standing slow-mode waves \citep{ofm02}. The properties of SUMER oscillations can be found in a review by \citet{wan11}. SDO/AIA as the most powerful solar EUV Imager so far, by virtue of consecutive observations with the wide field of view and broad temperature range, has captured longitudinal (standing and reflected propagating) waves in flaring coronal loops with the properties similar to those of SUMER oscillations \citep[e.g.][]{kum13,kum15,wan15}. The standing slow-mode waves were also recently found to be generated impulsively in fan-like coronal structures seen in the AIA 171 and 193 \AA\ bands \citep{pant17}. In addition, there is the increasing evidence suggesting that a kind of decaying quasi-periodic pulsations detected in solar and stellar flares could be associated with slow-mode oscillations \citep[see][for recent reviews]{vand2016,mclg2018}. This provides us a new avenue to explore the physical processes in stellar flares by the seismological techniques developed based on the solar observations \citep[e.g.][]{mit05,anf13,pugh16,real18}.

Remarkable theoretical attention has been given to the excitation, propagation, and damping mechanisms of observed slow-mode waves. The new observations in combination with theoretical progress in the understanding of these aspects have led to important breakthroughs in coronal seismology. For example, the signatures of slow-mode waves have been used to determine the plasma-$\beta$ and magnetic field in oscillating loops \citep{wan07,jess2016,nis17}, the polytropic index and transport coefficients \citep{vand11,wan15,wan19,krish18,krish19}, and constrain the coronal heating function \citep{nak17,real19,kolot19}. 

This review focuses on new observations of slow magnetoacoustic waves in flaring coronal loops and emphasizes their theoretical insights and seismological applications. Here, we review mainly published studies including the related supplemental material (see Figs.~\ref{fig:m3d} and \ref{fig:pulse_dur}), and provide some new calculations of non-ideal damping effects in typical hot coronal loops useful for the present discussion (see Table~\ref{tab:der} and Figs.~\ref{fig:tpd}, \ref{fig:tps}, and \ref{fig:ntl}). For readers interested in propagating slow-mode waves in coronal fan or plume structures, we refer to the review by \citet{ban20} in this issue. This article is organized as follows: we describe in Sect.~\ref{sct:tb} the solutions of standing slow-mode waves in a thin magnetic flux tube, demonstrate their observational characteristics by forward modeling, and review the motivations for studying the slow-mode waves with the hydrodynamic (HD) models. We review in Sect.~\ref{sct:swhlp} observations of standing and reflected propagating slow-mode waves in flaring coronal loops and in Sect.~\ref{sct:swflp} observations of standing slow-mode waves in warm coronal fan loops. We briefly compare the properties of slow-mode waves observed in solar and stellar flares in Sect.~\ref{sct:qpp}. We then review the wave excitation mechanism in Sect.~\ref{sst:exc} and the damping mechanism in Sect.~\ref{sst:dm}. We finally review some coronal-seismological applications in Sect.~\ref{sct:acs}, followed by conclusions and open questions in Sect.~\ref{sct:coq}.

\section{Theoretical basis}
\label{sct:tb}

\subsection{Standard cylinder model}
\label{sst:stm}
In the theoretical study of standing slow-mode wave, a magnetic cylinder model filled with uniform plasma is often used. Since a slow-mode wave is mainly driven by pressure gradient, its perturbations are dominated by the component along the magnetic field line. Often, this model uses ideal MHD equations and neglects non-ideal MHD terms and complexity in magnetic field configuration. 

We consider a standing slow-mode wave in a plasma embedded in a uniform magnetic flux tube. The magnetic field only has a component along the axis of the plasma cylinder (i.e., $z$-axis), $\bvec{B_0}=B_0 \bvec{\hat{z}}$. The equilibrium magnetic field $B_0$, plasma density $\rho_0$, and temperature $T_0$ are the piecewise functions of $r$-axis:
\begin{equation}
B_0,\rho_0,T_0=\left\{
\begin{array}{lr}
B_\i,\rho_\i,T_\i &: r \leq a, \\
B_\e,\rho_\e,T_\e  &: r>a,
\end{array}
\right.
\end{equation}
where $a$ is the radius of the flux tube. The subscripts \singlequote{$\i$} and \singlequote{$\e$} denote the internal and external values. 

The linearized ideal MHD equations \citep[e.g.][]{rud09,yuan15} give the perturbed variables that deviate from magnetostatic equilibrium: 
\begin{align}
\rho_1 &=-\div(\rho_0\bvec{\xi}), \label{eq:cont} \\
\rho_0\frac{\partial^2\bvec{\xi}}{\partial t^2}&=-\grad{P_{\T1}} + \frac{1}{\mu_0}[ (\bvec{B_0}\cdot\grad)\bvec{b_1}+(\bvec{b_1}\cdot\grad)\bvec{B_0}],\label{eq:moment}\\ 
\bvec{b_1}&=\curl(\bvec{\xi}\times\bvec{B_0}), \label{eq:mag}\\
p_1-C_\s^2\rho_1&=\bvec{\xi}\cdot(C_\s^2\grad{\rho_0}-\grad{p_0})  \label{eq:entropy},
\end{align}
where $\bvec{\xi}$ is the Lagrangian displacement vector, $p_0$ is the equilibrium plasma pressure, $\rho_1$, $p_1$ and $\bvec{b_1}$ are the perturbed plasma density, pressure, and magnetic field, $P_{\T1}=p_1+\bvec{b_1}\cdot\bvec{B_0}/\mu_0$ is the perturbed total pressure, $\mu_0$ is the magnetic permeability in free space. We define the key characteristic speeds to describe the loop system, $C_\s=\sqrt{\gamma p_0/\rho_0}$, $C_\A=B_0/\sqrt{\mu_0\rho_0}$, $C_\T=C_\A C_\s/\sqrt{C_\A^2+C_\s^2}$ are the acoustic, \alfven, and tube speed, respectively; $\omega_\s=C_\s k$, $\omega_\A=C_\A k$, $\omega_\T=C_\T k$ are the corresponding acoustic, \alfven, and tube frequencies, where $k=\pi n/L$ is the longitudinal wavenumber, $n$ is the longitudinal mode number ($n=1$ corresponds to the fundamental mode), $L$ is the loop length, $\gamma=5/3$ is the adiabatic index. In the piecewise-uniform equilibrium considered here the terms with gradients of the unperturbed plasma parameters are zero. 

Equations~(\ref{eq:cont})$-$(\ref{eq:entropy}) are solved in cylindrical coordinates ($r,\phi,z$). In the case of standing slow sausage mode (with the azimuthal wavenumber $m=0$), we analyze the perturbed quantities with Fourier decomposition. Considering the boundary condition at the footpoints for $v_\mathrm{z}$, i.e., $v_\mathrm{z}=0$ at $z=0$ and $z=L$, we assume the perturbed total pressure following a profile as  $P_{\T1}=A\RR(r) \cos (\omega t) \cos (k z)$, where $A$ is the amplitude of the perturbation and $\RR(r)$ is a dimensionless function depending on $r$. The perturbed thermodynamic quantities can obtained from Eqs.~(\ref{eq:cont})$-$(\ref{eq:entropy}) as,
\begin{align}
v_r &=-\frac{A}{\rho_0}\left(\frac{\d \RR}{\d r}\right)\frac{\omega}{\omega^2-\omega_\A^2}\sin (\omega t) \cos (k z), \label{eq:vr} \\
v_z &= \frac{A\RR C_\T^2 k \omega}{\rho_0C_\A^2(\omega^2-\omega_\T^2)}\sin (\omega t) \sin (k z),\\
v_\phi& =0, \\
\rho_1&=\frac{A\RR }{(C_\s^2+C_\A^2)}\frac{\omega^2}{(\omega^2-\omega_\T^2)} \cos (\omega t) \cos (k z), \\
T_1&=\frac{A\RR (\gamma-1)T_0}{\rho_0(C_\s^2+C_\A^2)}\frac{\omega^2}{(\omega^2-\omega_\T^2)} \cos (\omega t) \cos (k z), \label{eq:temp}
\end{align}
where $\RR(r)$ must satisfy the Bessel equation
\begin{equation}
\frac{\d^2\RR}{\d r^2}+\frac{1}{r}\frac{\d \RR}{\d r}-\kappa_r^2\RR=0, \label{eq:ptr}
\end{equation} 
where $\kappa_r^2=\frac{(\omega_\s^2-\omega^2)(\omega_\A^2-\omega^2)}{(\omega_\s^2+\omega_\A^2)(\omega_\T^2-\omega^2)}k^2$ is a modified radial wavenumber. \eqref{eq:ptr} can be derived from the linearized MHD equations by eliminating all the perturbed variables but $P_{T1}$ \citep[see][]{sak91}. Solving \eqref{eq:ptr} for both internal and external plasmas gives the solution
\begin{equation}
\RR= \left\{
\begin{array}{lr}
J_{0}(|\kappa_{ri}|r)  &: r \leq a, \\
K_{0}(\kappa_{re}r)  &: r>a,
\end{array}
\right.
\end{equation}
where $\kappa_{r\e}^2>0$ and $\kappa_{r\i}^2<0$, hence we re-define $|\kappa_{r\i}|=\sqrt{-\kappa_{r\i}^2}$. Considering the conditions for continuity of $\xi_r$ (or $v_r$ in the absence of steady flows along the loop) and $P_{T1}$ at the tube surface $r=a$ \citep{edw83,rud09}, the dispersion relation that determines the relation between the wavenumber $k$ and wave frequency $\omega$ can be derived, 
\begin{equation}
\frac{\kappa_{r\e}}{\rho_\e(\omega_{\A\e}^2-\omega^2)}\frac{K'_0(\kappa_{r\e} a)}{K_0(\kappa_{r\e}a)}=\frac{|\kappa_{r\i}|}{\rho_\i(\omega_{\A\i}^2-\omega^2)}\frac{J'_0(|\kappa_{r\i}| a)}{J_0(|\kappa_{r\i}| a)}, \label{eq:disp}
\end{equation}
where $J$ and $K$  are the Bessel function of the first kind  and modified Bessel function of the second kind. The subscript denotes the order and the prime sign represents the derivatives with respect to its independent variable. It is important to mention that for a flow tube (e.g., with shear flows in the solar wind) the appropriate boundary condition at the tube surface should be continuity of the normal displacement $\xi_r$ \citep{nak96}.

Equations~(\ref{eq:vr})$-$(\ref{eq:disp}) hold for both the fast and slow sausage modes. The solution corresponds to the slow sausage mode when the equations are solved in the acoustic frequency range $\omega _{\rm Ti}\leq \omega \leq \omega _{\rm si}$ \citep[e.g.][]{mor13, yuan15}, while the solution corresponds to the fast sausage mode when in the Alfv\'{e}n frequency range $\omega _{\rm Ai}\leq \omega \leq \omega _{\rm Ae}$\citep[e.g.][]{anto13, rezn14}. Since $v_z/v_r \propto (\omega ^{2}-\omega_{\rm A}^{2})/(\omega ^{2}-\omega_{\rm T} ^{2})$, it indicates that the slow sausage mode is dominated by the longitudinal motions ($v_z \gg v_r$), while the fast sausage mode is dominated by the radial motions ($v_r \gg v_z$). Readers who are interested in detailed discussion on coronal fast sausage modes are referred to a review by \citet{li20} in this issue. This property of the slow sausage mode allows one to model slow-mode waves in terms of infinite magnetic field for zero plasma-$\beta$ or thin flux tube approximations for non-zero magnetic effects \citep[e.g.][]{zhu96,nak00}. For the fundamental mode, since $\kappa_{r} a\ll{1}$, we have $\RR(r)>0$ and $\d \RR/\d r <0$ for $r\lesssim{a}$, thus the phase relationships between different perturbed quantities in time and space can be easily determined from Eqs.~(\ref{eq:vr})$-$(\ref{eq:temp}), which are important for our understanding of the observed wave features. The set of perturbed variables ($v_r$, $v_z$, $\rho_1$, $T_1$) affects the line emissions from coronal plasmas and is used by forward modeling to calculate the imaging and spectral emissions.

\begin{figure}
 	\centering
	\includegraphics[width=0.7\textwidth]{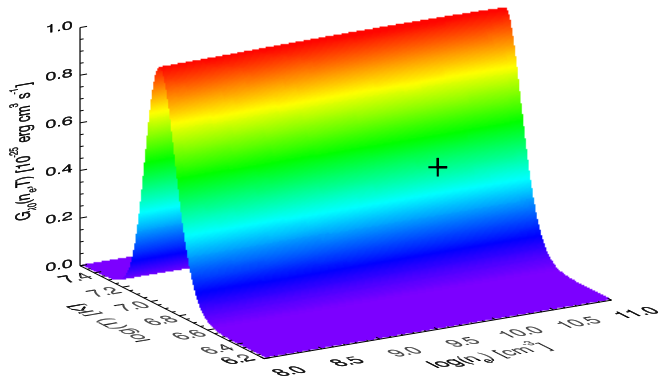}
	\caption{Contribution function $G(n_e,T)$ for the Fe\,{\sc xix} $\lambda1118.1$ line \citep[from][]{yuan15}. The peak formation temperature is $\log T = 6.95$ (or $T$=8.9 MK).}
	\label{fig:yuan2015_fig7}   
\end{figure}

\subsection{Forward modeling}
\label{sst:fm}
Forward modeling uses numerical models of magnetized plasma, calculates the plasma emissions in various electromagnetic bandpasses, and synthesizes the imaging or spectrographic signals expected in space-borne or ground-based instruments. Early attempt was done to model the observational signature of standing slow-mode wave in a one-dimensional hydrodynamic loop \citep{tar07}. A full three-dimensional forward modeling of standing slow-mode wave in a coronal loop was done by \citet{yuan15}. 

\citet{yuan15} modeled the imaging and spectral observational features of a hot coronal loop for a standing slow-mode wave using the standard model as described in Section~\ref{sst:stm}. The loop's temperature was set to $6.4\unit{MK}$, at which iron atoms are ionized to Fe\,{\sc xix} and have strong line emission centered at 1118.1 \AA. This emission line was covered by the SUMER spectrograph, and was used intensively to study the standing slow-mode waves in hot coronal loops \citep{wan11}. 

A contribution function $G(n_e,T)$ that contains the terms related to atomic physics \citep{land13}, is drawn for the Fe\,{\sc xix} $\lambda1118.1$ line in \figref{fig:yuan2015_fig7}. We can see that the contribution function is a weak function of plasma density, however it is strongly dependent on plasma temperature. The peak formation temperature is about 8.9 MK in the present case. Below the peak temperature, the contribution function increases with rising temperature; above the peak temperature the contribution function decreases with increasing temperature. This feature is extremely important as in hot flaring region, heating and cooling are not usually in balance with each other, and could cause sharp temperature variations in a coronal loop. Thus, one has to be very cautious in interpreting the observations, by taking into account the strong temperature dependence of the emission.   

\begin{figure}
\centering
\includegraphics[width=0.9\textwidth]{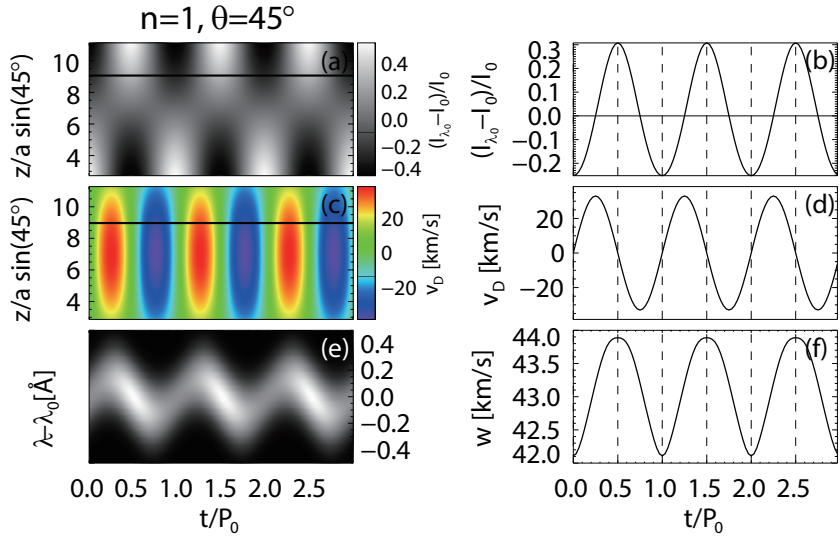}
\caption{(a) and (c) Time distance plots of the relative intensity variation and  $v_z$ along the central axis of the loop. Time variations of (b) intensity, (d) Doppler velocity, (e) spectral profile, and (f) line width at the marked location (horizontal lines in (a) and (c)). All the information is extracted at a viewing angle of $\theta=45^\circ$. Adapted from \citet{yuan15}.}
\label{fig:yuan2015_fig3}
\end{figure}

\figref{fig:yuan2015_fig3} presents the observational feature of a fundamental standing slow-mode wave in a coronal loop, obtained from forward modeling \citep{yuan15}.  In \figref{fig:yuan2015_fig3}a, we  could see that the intensity varies periodically, it has large amplitude at the footpoints, the oscillation amplitude approaches zero at the loop apex, indicative of a node structure of a standing wave. At two footpoints, the intensities oscillate in anti-phase with each other, that is the pattern of two antinode structures. \figref{fig:yuan2015_fig3}c shows the variation of Doppler velocity, it clearly shows the pattern of a standing wave: an antinode with the strong amplitude measured at the loop apex, the amplitude approaches zero at the footpoints where the fixed node for $v_z$ is located. The intensity and Doppler velocity oscillate with a $90\deg$ phase shift (Figs.~\ref{fig:yuan2015_fig3}b and d), this feature was first observed with SUMER \citep{wan03b}. A slow-mode standing wave also exhibits line width variations with the same periodicity (Fig.~\ref{fig:yuan2015_fig3}f), but the modulation depth is not very strong, no observation that confirms this effect has been reported yet. 

\begin{figure}
\centering
\includegraphics[width=0.9\textwidth]{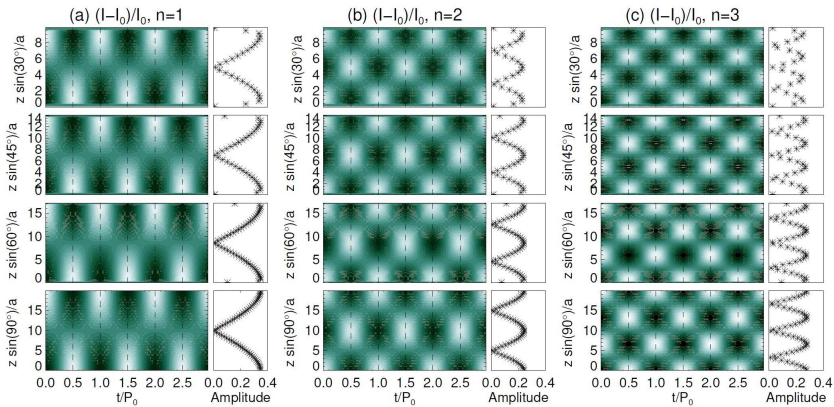}
\caption{Time-distance plots of synthetic AIA 94 \AA{} emission intensity along the central axis of the loop for the longitudinal mode numbers (a) $n = 1$, (b) $n=2$, and (c) $n=3$. The spatial profiles of amplitude (curves) indicate the nodal structures. Adapted from \citet{yuan15}. }
\label{fig:yuan2015_fig8}
\end{figure}

\citet{yuan15} predicted the observational features of a standing slow-mode wave in an imaging instrument, such as SDO/AIA. \figref{fig:yuan2015_fig8} presents the predicted features for the longitudinal modes $n=1$, 2, and 3. One could see that the main identifiable feature is the nodal structure in the oscillation amplitude along a coronal loop, the number of nodes determines the longitudinal mode number. The intensity oscillations at two points across a node are in anti-phase with each other. \citet{wan15} observed the excitation and formation of standing slow-mode waves in closed flaring coronal loops. The associated linear wave theory shows that the thermal conduction coefficient is suppressed by at least a factor of 3 in the hot flare loop at 9 MK and above, whereas the rapid damping indicates that the classical compressive viscosity coefficient needs to be enhanced by a factor of up to  15 in practice. \citet{pant17} observed standing slow-mode waves in diffuse fan-like coronal structures excited by a global EUV wave. All observational patterns (see Sects.~\ref{sct:swhlp} and \ref{sct:swflp} in detail) match very well with this prediction \citep{yuan15}. It is still an open question how a standing slow-mode wave is formed within the fan-like coronal structures, in which any individual magnetic thread could be rooted to an opposite polarity on the solar disk or extend radially into the heliosphere. 

\subsection{Modeling of slow-mode waves in an infinite magnetic field approximation}
\label{sst:mswa}

Long-period oscillations are frequently and mostly detected in the light curves of flaring loops, but also of other transient loop events. This suggests that these oscillations are directly connected to the impulsive nature of the events. Indeed, we expect that a sudden and short heat burst triggers all kinds of propagating disturbances and a magnetic loop acts as an almost ideal closed waveguide, where the disturbance travels back and forth becoming a standing wave. The modulation of the light curves, which could be of high amplitude, indicates density perturbations, and therefore that the oscillations are most probably connected to the propagation of magnetoacoustic waves. The long period further restricts the parameter space to slow modes. As such, these long-period oscillations could provide important evidence for the connection between slow-mode waves and impulsive heating in closed magnetic loops, and are potentially useful for the diagnostic of the heating processes in solar and stellar flares by coronal seismology \citep{wan11}.

Since there is no way at the moment to detect a direct signature of coronal heating function, the study of this connection requires theoretical and modeling efforts. A coronal loop is typically considered as a magnetic flux tube extending between two distant footpoints anchored in the photosphere. The strong magnetic field confines the plasma inside the loops. A heat burst is expected to perturb the whole magnetic system, and to trigger proper MHD waves, involving both the plasma and the magnetic field. However, the corona is a strongly magnetized environment, and the plasma-$\beta$ is typically very low \citep[e.g.][]{Gary2001a}. Even during intense flaring events, it is observed that the flaring loop geometry often does not change considerably \citep{Pallavicini1977a}. Thus, although in principle proper MHD disturbances, such as sausage modes \citep{tian16,nak18}, may be important, it is not unreasonable to assume that the magnetic field holds ``rigid" (the so-called infinite magnetic field approximation). We can therefore consider a coronal loop as a single solid tube and focus our attention to plasma as a confined fluid, where only acoustic modes propagate along the magnetic field. We can describe the wave propagation with pure hydrodynamics for a compressible fluid. The simple loop geometry allows us to describe the fluid with a single coordinate following the curved field lines. The very low conductivity across the field lines allows to assume that the energy is also transported exclusively along the field lines. In spite of these simplifications, a realistic description must include many physical effects, often highly nonlinear, such as the gravity component and the nonlinear plasma thermal conduction along the field line, radiative losses from optically thin fully (or partially)-ionized medium, and an external energy input to account for coronal heating. For highly transient events, this energy input is strong and bursty and the plasma response is nonlinear in time as well. The related time-dependent hydrodynamic equations are quite complex.

In addition, impulsive events necessarily imply a strong mass exchange with the low and dense atmosphere layers, because the sudden pressure increase drives a strong mass flow upwards from the chromosphere. A realistic investigation must therefore describe an atmosphere including both the corona and the chromosphere (at least) linked through the steep transition region. All these ingredients make the system mathematically complex to describe fully self-consistently. Although the study of the slow-mode wave evolution by linearizations is possible \citep[e.g.][]{alg14,bah18}, they provide partial answers, and for an accurate investigation of the initial impulsive evolution and for a proper comparison with observations the full physical description is necessary, which is only possible with numerical modeling at the moment. The slow-mode waves have been intensively studied using the 1D HD models to understand their observed features in the excitation (see Sect.~\ref{sst:exc}) and dissipation processes (see Sect.~\ref{sst:dm}), as well as to constrain on the heating function in observed flaring coronal loops (see Sect.~\ref{sst:hf}).

\section{Observations of slow-mode waves in hot flaring loops}
\label{sct:swhlp}
\subsection{Standing and reflected propagating modes}
\begin{figure}
         \centering
         \includegraphics[width=0.8\textwidth]{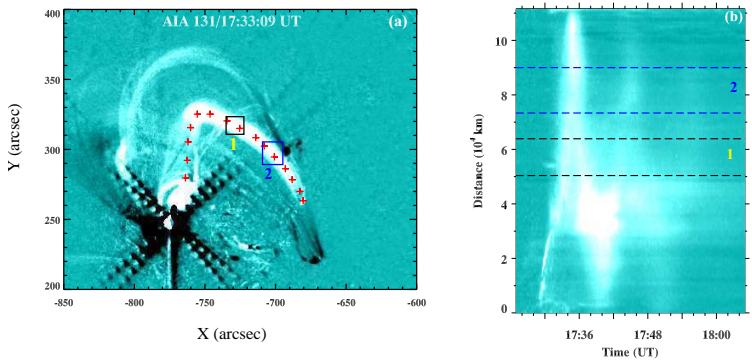}
         \caption{Longitudinal intensity oscillations discovered with SDO/AIA \citep[from][]{kum13}. (a) AIA 131 \AA\ base difference image of the oscillatory hot loop (marked with pluses). (b) Time-distance plot of the intensity for the selected path along the loop.}
         \label{fig:lio}
 \end{figure}

Slow magnetoacoustic oscillations of hot ($>$6 MK) coronal loops were first discovered with the imaging spectrometer, SOHO/SUMER, in flare emission lines (mainly, Fe\,{\sc xix} and Fe\,{\sc xxi}) as periodic variations of the Doppler shift \citep[see][]{wan11}. Similar Doppler shift oscillations were also detected in the flare emission lines, S\,{\sc xv} and Ca\,{\sc xix}, with Yohkoh/BCS \citep{mari05,mari06}. These oscillations are mostly interpreted as the fundamental standing slow-mode waves because their periods correspond to twice the acoustic travel time along the loop and there is a quarter-period phase lag between velocity and intensity (that mainly relates to density) disturbances detected in some cases \citep[e.g.][]{wan02,wan03a,wan03b}. \citet{kum13} first reported the detection of longitudinal intensity oscillations in flaring loops with SDO/AIA in high-temperature channels, namely 94 \AA\ (7 MK) and 131 \AA\ (11 MK) (see Fig.~\ref{fig:lio}). These oscillations, shown with the properties (such as periods and decay times) matching the SUMER oscillations, have been interpreted as either a reflected propagating slow-mode wave  \citep{kum13,kum15,mand16,nis17}, or a standing slow-mode wave \citep{wan15}.

\begin{figure}
         \centering
         \includegraphics[width=0.8\textwidth]{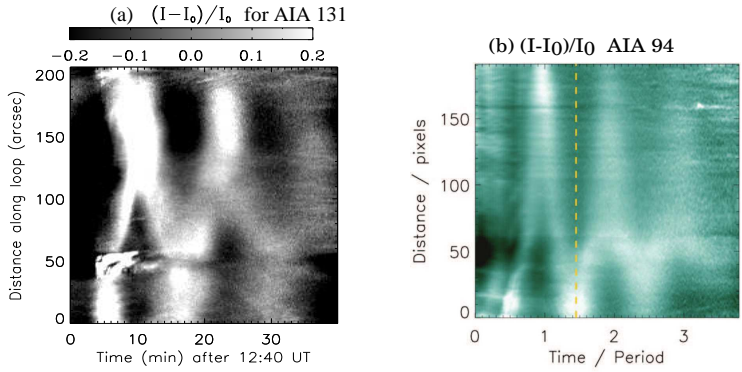}
         \caption{Time-distance diagrams showing longitudinal intensity oscillations in a hot flaring loop. (a) The event showing the standing pattern observed in SDO/AIA 131 \AA\ on 2013 December 28 \citep[from][]{wan15}. (b) The event showing the reflective feature observed in AIA 94 \AA\ on 2012 May 7 \citep[from][]{nak19}.}
         \label{fig:srp}
 \end{figure}

Figure~\ref{fig:srp} demonstrates that the two modes can be distinguished based on their spatiotemporal features in intensity oscillations. The fundamental standing wave shows the antiphase oscillations between the two legs (see panel (a)), whereas the reflected propagating wave exhibits a ``zigzag" pattern (see panel (b)) and the propagating speeds (that can be estimated by the slope of the bright ridges) are close to the speed of sound as determined from the loop temperature \citep[e.g.][]{mand16,wan18}. In addition, \citet{mand16} reported the detection of a number of reflected longitudinal wave events in hot loops with Hinode/XRT. These intensity oscillations decay rapidly as the perturbation moves along the loop and eventually vanishes after one or more reflections. Observations from both SDO/AIA and Hinode/XRT have confirmed that longitudinal oscillations are produced by a small (or micro-) flare at one of the loop's footpoints, which was also suggested as a trigger for the SUMER loop oscillations \citep{wan05}. However, physical conditions responsible for the formation of a fundamental standing wave or a reflected propagating wave by the footpoint heating are still poorly understood. An overview of theoretical studies on the wave excitation mechanism based on MHD simulations is given in Section~\ref{sst:exc}.

In addition, quasi-periodic pulsations in the emission from brightening regions of hot loop systems were detected with SDO/AIA in high-temperature channels, and the duration and location of the heat pulses producing them were investigated based on the HD modeling \citep[][see also Sect.~\ref{sst:hf}]{real19}.

\subsection{Physical properties}
\label{sst:spp}
From statistical analysis of 54 oscillations in 27 flare-like events observed with SUMER, \citet{wan03a} obtained oscillation periods $P=7-31$ min with a mean of $17.6\pm5.4$ min, and decay times $\tau=6-37$ min with a mean of $24.6\pm7.0$ min. For seven SUMER oscillation events associated with Yohkoh/SXT observations, \citet{wan07} estimated the temperature of hot loops $T=6.1-7.0$ MK with a mean of $6.6\pm0.4$ MK and the electron density $n=(3.4-12)\times10^9$ cm$^{-3}$ with a mean of $(7.4\pm3.3)\times10^9$ cm$^{-3}$ using the filter ratio method. The measured loop temperature from soft X-ray (SXR) observations is consistent with the fact that the SUMER oscillations are mostly detected in flare emission lines with the peak formation temperature higher than 6 MK. \citet{mari06} analyzed 20 flares showing the Doppler shift oscillations with Yohkoh/BCS spectra and found average oscillation periods of $5.5\pm2.7$ min and decay times of $5.0\pm2.5$ min. Note that the BCS oscillations are detected in the hotter flare lines at 12$-$14 MK. If we estimate the theoretical oscillation period using $P\approx2L/jC_s$, where $L$ is the loop length, $C_s=152\sqrt{T({\rm MK})}$ km~s$^{-1}$ is the adiabatic sound speed at the loop average temperature, and $j$ is the harmonic number of the standing mode, we obtain the following relations
\begin{eqnarray}
\frac{L_{\rm sum}}{L_{\rm bcs}}&=&\frac{P_{\rm sum}}{P_{\rm bcs}}\frac{C_s^{\rm sum}}{C_s^{\rm bcs}}=\frac{P_{\rm sum}}{P_{\rm bcs}}\left(\frac{T_{\rm sum}}{T_{\rm bcs}}\right)^{1/2} ~~~({\rm if} j_{\rm sum}=j_{\rm bcs}), \\
\frac{j_{\rm bcs}}{j_{\rm sum}}&=&\frac{P_{\rm sum}}{P_{\rm bcs}}\frac{C_s^{\rm sum}}{C_s^{\rm bcs}}=\frac{P_{\rm sum}}{P_{\rm bcs}}\left(\frac{T_{\rm sum}}{T_{\rm bcs}}\right)^{1/2} ~~~({\rm if} L_{\rm sum}=L_{\rm bcs}).
\end{eqnarray}
From measurements of the physical parameters above, we found that $L_{\rm sum}/L_{\rm bcs}$=2.3 if assuming the fundamental modes for both SUMER and BCS oscillations, while $j_{\rm bcs}/j_{\rm sum}$=2.3 if assuming that SUMER and BCS observed the loops of a similar size. This theoretical estimation suggests that one reason for $P_{\rm bcs}<P_{\rm sum}$ (besides their difference in the observed temperatures) could be that the oscillating loops detected by BCS are systematically shorter than those by SUMER if they are oscillating with the same longitudinal harmonic, or this may instead suggest that the SUMER oscillations are in the fundamental mode of oscillation while the BCS oscillations in the second harmonic. Since the BCS viewed the entire Sun the fundamental mode oscillations are preferentially detected in the loops near the limb, while the second harmonics are preferentially detected on the disk due to projection effects. This is because the fundamental modes have an antinode in velocity at the loop apex, while the second harmonics have the antinodes in velocity at the loop legs. \citet{mari06} identified that 18 out of 20 flares showing the oscillations were located near the limb, thus it indirectly indicates that the BCS oscillations are most likely in the fundamental mode as the SUMER oscillations. This conclusion is also supported by the result from a case analysis in \citet{mari06}.

 \begin{figure}
 \centering
          \includegraphics[width=0.9\textwidth]{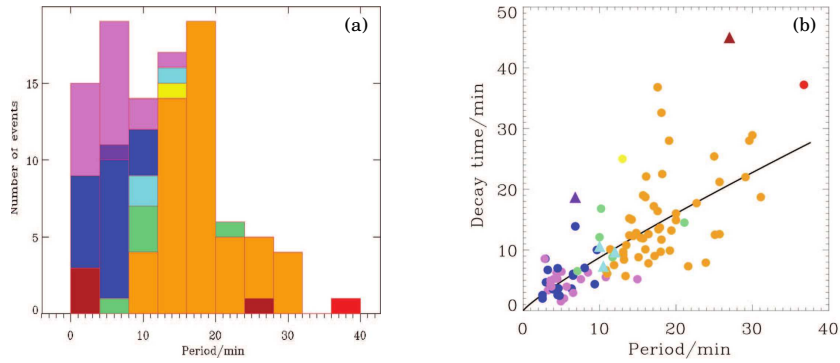}
         \caption{Physical properties of observed slow-mode oscillations \citep[from][]{nak19}. (a) Distribution of the wave periods for different detection temperatures. (b) Scaling of the damping time with the period. In (a) and (b) the colors represent different temperatures: violet for 14 MK, magenta for 13 MK, blue for 12 MK, cyan to 10 MK, green for 8.9 MK, yellow for 7 MK, orange to 6.3 MK, red to 2 MK, and brick for 0.6 MK. In (b) the circles show SUMER and BCS oscillations and triangles show AIA oscillations. The black line indicates the best-fitting power law.}
         \label{fig:wpf}
 \end{figure}

For the four AIA longitudinal oscillations detected in the high-temperature channels (3 associated with the GOES C-class flares and 1 with the B-class flare) reported in the literature \citep{kum13,kum15,wan15,nis17}, we estimate their average physical parameters to be $P=9.8\pm2.2$ min, $\tau=11.9\pm6.0$ min, $T=9.2\pm1.3$ MK, $n=(5.7\pm3.0)\times10^9$ cm$^{-3}$, and $L=145\pm30$ Mm. They are well in line with those for the SUMER oscillations. Figure~\ref{fig:wpf}a shows the distribution of the periods for slow-mode oscillations detected with various instruments, where different colors represent the temperatures of the emission lines or bandpasses used in the detections. The temperature distribution indicates that the detected oscillations in the hotter channels have systematically shorter periods. Figure~\ref{fig:wpf}b shows that the scaling between the damping time and the oscillation period can be roughly fitted with a power-law relation $\tau=aP^b$. \citet{nak19} obtained $a=1.18\pm0.4$ and $b=0.87\pm0.1$ for longitudinal oscillations observed from SUMER, BCS, and AIA. This result is very close to the power law ($\tau=1.30P^{0.81}$) obtained by fitting to the combined SUMER and BCS data \citep{mari06} and to the power law ($\tau=0.84P^{0.96\pm0.18}$) obtained by fitting the improved measurements of SUMER oscillations with a correction of the effects of the flows \citep{wan05}. The nearly linear scaling relationship between $\tau$ and $P$ can be interpreted based on a linear theory of slow-mode waves damping due to non-ideal MHD effects (see Sect.~\ref{sss:nme}).

Recently, \citet{cho16} found the damped harmonic oscillations in 42 flares detected in the hard X-ray emission with RHESSI in the energy band 3$-$25 keV, showing $P=0.90\pm0.56$ min and $\tau=1.53\pm1.10$ min with a mean ratio $\tau/P=1.74\pm0.77$ and a power-law scaling fit $\tau =(1.59\pm1.07)P^{0.96\pm0.10}$. They interpreted these oscillations as resulting from standing slow modes or kink modes in flaring loops because the obtained power index of nearly 1 is consistent with that of the scalings both for longitudinal oscillations observed with SUMER and transverse oscillations observed with TRACE or AIA \citep{ver13,god16,nech19}. However, the former has much higher likelihood than the latter as the transverse oscillations are rarely observed in hot flaring loops and they are typically associated with weak intensity variations \citep[e.g.][]{coop03,whi12a,whi12b}. In addition, we notice a distinct difference in periods between RHESSI and SUMER detected oscillations. This difference could be explained by the high energy band of RHESSI that tends to detect much hotter and shorter loops than those detected in lower energy emission by SUMER instrument. For example, for typical hotter ($T\sim25$ MK) and shorter ($L\sim20$ Mm) flare loops that are sensitive to RHESSI \citep{jiang06,casp14,ryan14}, the fundamental slow modes have the expected periods $P\approx2L/C_s=0.88$ min, agreeing well with the measured periods in \citet{cho16}.

 \begin{figure}
         \centering
         \includegraphics[width=0.8\textwidth]{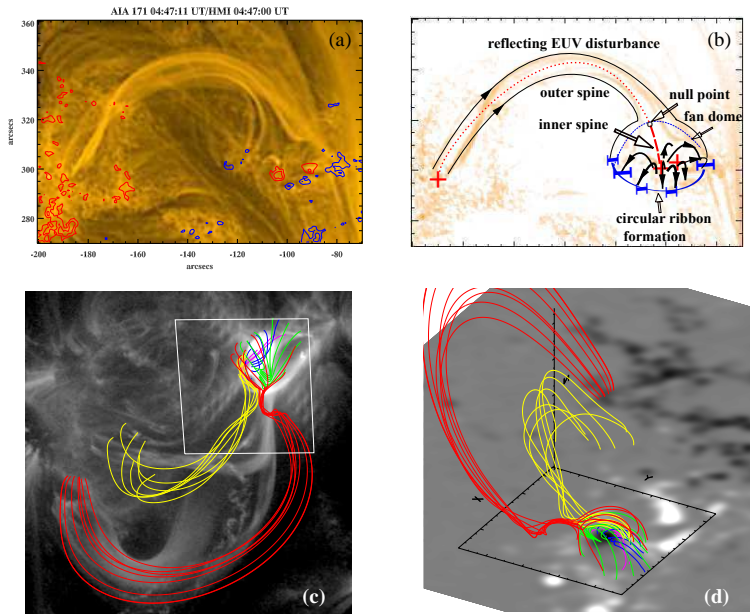}
         \caption{Fan-spine magnetic topology associated with the trigger of longitudinal oscillations. (a) Cooling coronal loops in AIA 171 \AA\ during the postflare phase showing the fan-spine topology. (b) The schematic cartoon of the fan-spine topology. (c) AIA 131 \AA\ image, overlaid with magnetic field lines calculated using a nonlinear force-free field model. (d) Side view of the magnetic skeleton, superposed on an HMI radial field map (with smoothing and scaled between $\pm$1100 G). Panels (a) and (b) from \citet{kum15} and Panels (c) and (d) from \citet{wan18}.}
         \label{fig:mtt}
 \end{figure}

In addition, the SUMER oscillations often have large amplitudes with respect to the sound speed with the relative amplitude $A=V_D/C_s>20\%$, where $V_D$ is the Doppler velocity amplitude of a fitted oscillation \citep{wan03a,nak19}. \citet{verw08} found the linear scaling relationship $\tau/P\propto(0.34\pm0.04)A^{-1}$ for the SUMER data, and \citet{nak19} obtained $\tau/P\propto0.56 A^{-0.33}$ for the combined SUMER and BCS data. The dependence of the damping time on the oscillation amplitude indicates the nonlinear nature of the damping. \citet{nak19} further suggested that the reflective feature of longitudinal oscillations observed with AIA and XRT could be related to the competition between the nonlinear and dissipative effects.

\subsection{Magnetic topology for wave trigger}
High-resolution EUV observations from SDO/AIA have revealed that the trigger of fundamental standing and reflected propagating slow-mode waves are commonly associated with small circular-ribbon flares at one footpoint of a coronal loop \citep[e.g.][]{kum15,wan18}. Figure~\ref{fig:mtt} demonstrates two examples. The emission features and magnetic field extrapolation suggest that the circular-ribbon flares are caused by magnetic reconnections at a coronal null point in a fan-spine magnetic topology \citep[e.g.][]{mass09,wang12,sun13}. The impulsive magnetic energy release heats the large spine loop and generates a slow-mode wave which then reflects back and forth in the heated loop, ultimately forming a standing wave. Hot and cool plasma ejections with speeds on the order of 100$-$300 km~s$^{-1}$ are often found to be associated with the initiation of such flares \citep{kum13,kum15,mand16,nis17}. The impulsive flows could be evidence for a mini-filament (or small flux rope) eruption that triggers the flare and associated waves \citep[e.g.][]{sun13,wyp17}. It was also found that a 1600 \AA\ brightening appears at the remote footpoint location before the arrival of the main hot plasma disturbance from the flare site \citep{wan18}. This may indicate that the loop is heated by energetic particles or heat flux from the reconnection region. Since magnetic structure of the fan-spine magnetic topology is relatively stable (compared to the flare and wave timescales) it allows the recurrence of non-eruptive (or confined) flares. As such, this topology can also explain the trigger of SUMER oscillations, which were observed frequently recurring in the same loop system \citep{wan11}. The SUMER oscillations have another distinct feature that they often started with high-speed hot flows \citep{wan05}. It also supports this scenario.

\begin{figure}
        \centering
 \includegraphics[scale=0.7,angle=90]{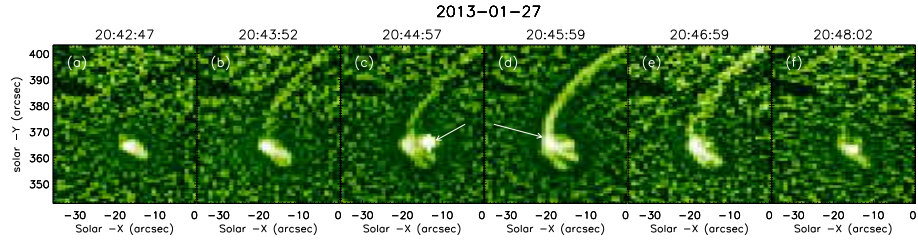}
\caption{Evolution of a coronal loop's footpoint recorded with the Hinode/XRT Be-thin filer. Brightening at the possible emerging bipolar structure is labelled by arrows. From \citet{mand16}.}
\label{fig:mandal2016_fig12}
\end{figure}

In addition, by tracing several wave events simultaneously observed with Hinode/XRT and SDO/AIA, \citet{mand16} suggested that reflective propagating slow-mode waves could be triggered by small-scale energy releases by magnetic reconnection, such as microflares and coronal jets. Figure~\ref{fig:mandal2016_fig12} shows an example of such events seen in SXR with XRT. Such microflares and jets have been interpreted by the breakout model of solar eruptions in a fan-spine magnetic topology based on the 3D MHD simulations \citep{wyp17,wyp18} and recent AIA observations \citep{kum18,kum19}. We infer that when a microflare happens in this type of topology of a closed outer spine, because the local ``magnetic breakout" is not strong enough to disrupt the entire loop system, the ejected hot plasma and associated pressure disturbances are confined in the loop forming a reflected propagating slow-mode wave.

\section{Observations of standing slow-mode waves in coronal fan loops}
\label{sct:swflp}
\begin{figure}
         \centering
         \includegraphics[width=0.9\textwidth]{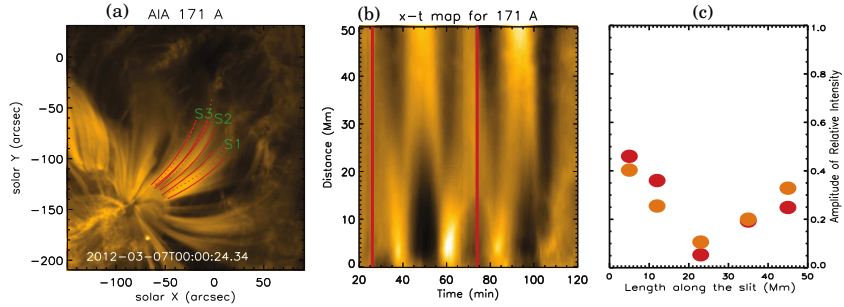}
         \caption{Observations of standing slow-mode waves in coronal fan loops \citep[from][]{pant17}. (a) AIA 171 \AA\ image. (b) Time-distance map corresponding to a curved slice $S1$ marked in (a). (c) Spatial variation of the relative amplitude of intensity oscillations for $S1$. Two vertical red lines in (b) represent the time when two blast waves impacted the fan loop system. }
         \label{fig:ssf}
 \end{figure}

Recently, standing slow-mode waves were discovered in warm coronal fan loops with SDO/AIA \citep{pant17}. These longitudinal oscillations were triggered by global EUV waves that originated from a distant AR due to X-class flares. The intensity oscillations were visible in both the 171 and 193 \AA\ channels but more evident in 171 \AA\ (see Fig.~\ref{fig:ssf}). The oscillation period was estimated to be $\sim$28 min in 171 \AA, slightly longer than that ($\sim$23 min) in 193 \AA. For the projected loop length $L\sim63$ Mm, the phase speed estimated using $V_p=2L/P$ was 75 and 92 km~s$^{-1}$ for the 171 and 193 \AA\ channels, respectively. The measured phase speed and its temperature-dependent behavior are consistent with the interpretation of observed intensity oscillations as slow-mode waves \citep{kris12,urit13}. Furthermore, the spatial features such as the antiphase oscillations between two footpoints (Fig.~\ref{fig:ssf}b) and the presence of a node in the middle of the loop (Fig.~\ref{fig:ssf}c) suggest that they are likely the fundamental standing mode. The standing slow-mode waves in the fan loops show a weak decay, compared to those in hot flaring loops. It could be because the fan loops are relatively cool ($\sim$0.7 MK) and the oscillations have longer periods. In such a condition thermal conduction as a dominant damping mechanism for slow-mode waves becomes less efficient (see Eq.~\ref{equ:wtt} in Sect.~\ref{sss:nme}). In addition, it is worth mentioning that only one footpoint of the fan loops is clearly visible. As such, \citet{pant17} suggested two scenarios to explain the possible reflection of the wave from the other end of the loop. One scenario is that there is an antinode of the oscillations at the other footpoint, but its signature is not obvious because the fan loops are divergent towards the other end. The other scenario is that the antinode could be present at the region of large density gradient close to the other end of the fan loop. So far there are no modeling studies to address the excitation mechanism of standing slow-mode waves in warm fan loops.

\section{Comparison of slow-mode waves observed in solar and stellar flares}
\label{sct:qpp}

Quasi-periodic pulsations (QPPs), characterized by time variations in the light curves of the flare emission are common features of solar and stellar flares (see e.g., the review by Zimovets et al. 2020, in this issue). The origin of QPPs could be related to oscillatory reconnection \citep[e.g.][]{ofm06,kup2020} and MHD oscillations \citep[e.g.][]{nak04,nak09,mclg2018}. Here we emphasize a kind of QPPs showing damped harmonic-type oscillations in the decay phase of flares, as it was proposed by \citet{nak2019}. Both solar and stellar observations have suggested that these kind of QPPs are most likely caused by standing slow-mode oscillations in hot flare loops \citep[e.g.][]{cho16,kup19}, including the decaying 5-min QPP in the most powerful solar flare of Cycle 24 with the energy in the realm of stellar flares \citep{kolot18}. For solar flares the interpretation of such QPPs, detected in the total X-ray flux over the full disk such as obtained from GOES and RHESSI, may resort to the associated imaging observations (e.g., from SDO/AIA and Hinode/XRT), from which the directly measured flaring loop length can be used to identify the wave modes \citep{kim12,kum15,kup19}. For stellar flares, however, which are spatially-unresolved, the loop length needs to be constrained from other information (independent of any oscillation) such as the temporal shape and thermal properties of the flare by an analogy with solar flare loop models \citep[e.g.][]{fav05,pan08}. In addition, the decaying QPPs have also been interpreted based on the hydrodynamic loop modeling with impulsive heating (see Sect.~\ref{sst:hf}).

\begin{table*}
\begin{threeparttable}
\caption{Observations of stellar flares showing the decaying harmonic oscillations, which were interpreted due to standing slow magnetoacoustic waves\tnote{a}}
\label{tab:qpp}       
\begin{tabular}{lccccll}
\hline\noalign{\smallskip}
Study  & $P$ (min) & $\tau$ (min) & $N$ & Wavelength & Instrument & Phase of flares \\
\noalign{\smallskip}
\hline\noalign{\smallskip}
\citet{mit05} & 12.5 & 33.3 & 1 & SXR & XMM-Newton & flat-top peak \\
\citet{sri13}\tnote{b} & 21 & 47 & 1 & SXR & XMM-Newton & decay phase \\
              & 11.5 & 47 & -- & -- & -- & -- \\
\citet{cho16} & 16.2$\pm$15.9 & 27.2$\pm$28.7 & 36 & SXR & XMM-Newton & decay phase \\
\citet{real18} & 167$\pm$17 & -- & 2 & SXR & Chandra & peak + decay \\
\citet{wel06} & 0.50$-$0.67 &  -- & 4 & UV & GALEX & rising + decay \\
\citet{anf13} & 32 & 46 & 1 & WL & APO & decay phase \\
\citet{bal15} & 8.2$\pm$3.6 & -- & 7 &  WL & Kepler & decay phase \\
\citet{pugh15}\tnote{b} & 78$\pm$12 & 80$\pm$12 & 1 & WL & Kepler & decay phase \\
              & 32$\pm$2 & 77$\pm$29 & -- & -- & -- & -- \\
\citet{pugh16} & 37.4$\pm$21.6 & 41.5$\pm$35.8 & 11 & WL & Kepler & decay phase \\
\noalign{\smallskip}\hline
\end{tabular}
  \begin{tablenotes}
  \item[a] In column names $P$ represents the oscillation period, $\tau$ the decay time, and $N$ the number of analyzed events. In the 3rd column, the item with `--' means that the decay time was not measured in the referenced study. In the 5th column, WL means the white light.
  \item[b] The oscillations show the multiple periodicities.
  \end{tablenotes}
\end{threeparttable}
\end{table*}

Table~\ref{tab:qpp} lists some studies on the stellar QPPs with a rapidly decaying harmonic feature in accord with standing slow-mode waves. It shows that the timescales of oscillations are distributed over a broad range of periods ($P$=30~s $-$ 3~hrs). If these QPPs are caused by standing slow-mode waves, this could imply the variety of the length scales in the stellar flare loops (e.g., $L\approx20-2000$ Mm if the plasma temperature at $T=5-50$ MK and for a fundamental mode). Recent statistical studies showed that the decay times and periods for such stellar QPPs follow approximately a linear relationship (i.e., $\tau\propto{P}$) \citep{pugh16,cho16}. This scaling agrees well with that for the slow-mode oscillations detected in solar flaring loops (see Sect.~\ref{sst:spp}). Another obvious feature for the decaying QPPs in stellar flares is that they are often detected in the white-light emission of the cooler photospheric/chromospheric plasma \citep[e.g.][]{bal15,pugh16}, while those in solar flares are mainly detected in the SXR and EUV emissions of the hot flaring plasma \citep[e.g.][]{cho16,nak19}. Motivated by multi-wavelength observations of solar flares, a scenario has been suggested to explain the origin of white-light QPPs by non-thermal electrons periodically precipitating into the lower layers of the stellar atmosphere due to the periodically induced magnetic reconnection by longitudinal waves \citep[for a detailed discussion, see][]{anf13}.

\section{Excitation mechanisms}
\label{sst:exc}
\subsection{Modeling of standing waves}
\label{sst:msw}
Modeling of standing slow magnetoacoustic waves in coronal AR loops were performed in many studies in the past in order to understand the damping and excitation mechanisms \citep[see the reviews by][]{wan11,wan16}. \citet{ofm02} developed the first nonlinear 1D MHD model with thermal conduction and viscosity to study the damping of nonlinear slow-mode waves in hot coronal loops observed with SOHO/SUMER. \citet{tar05,tar07} studied the excitation of the SUMER oscillations using the field-aligned 1D loop model extended by including gravity and inhomogeneous atmosphere such as temperature and density stratifications. Using a similar model, \citet{mend06} showed that random energy release near either one or both footpoints of the loop can produce intermittent patterns of the standing waves due to interference. Their simulations suggest a possible excitation mechanism for weakly-damped or undamped standing slow-mode waves observed with Hinode/EIS in warm (1$-$2 MK) coronal loops \citep[e.g.][]{erdt08,mar08,mar10}. In addition, the EIS-observed non-decaying oscillations may also be explained by the wave-caused misbalance between heating and cooling processes in the corona (see \citealt{kolot19} and Sect.~\ref{sss:whc}). Recently by considering viscous and thermal conduction damping with application to SDO/AIA observations of standing slow-mode waves in a hot flaring loop, \citet{wan18} and \citet{wan19} found that the anomalously enhanced viscosity may play an important role in wave excitation and quick damping. Two-dimensional MHD models of standing slow magnetoacoustic waves were developed in coronal arcade loops \citep[e.g.][]{selw07,ogr09,gru11}, motivated by SUMER and later TRACE high-resolution EUV observations. Impulsively generated slow standing waves in 3D corona loop structures were modeled in the past with various excitation methods such as fast mode waves, pressure pulses and flows \citep{selw09,pas09,ofm12}.

\begin{figure}
         \centering
         \includegraphics[width=0.9\textwidth,height=0.4\textheight]{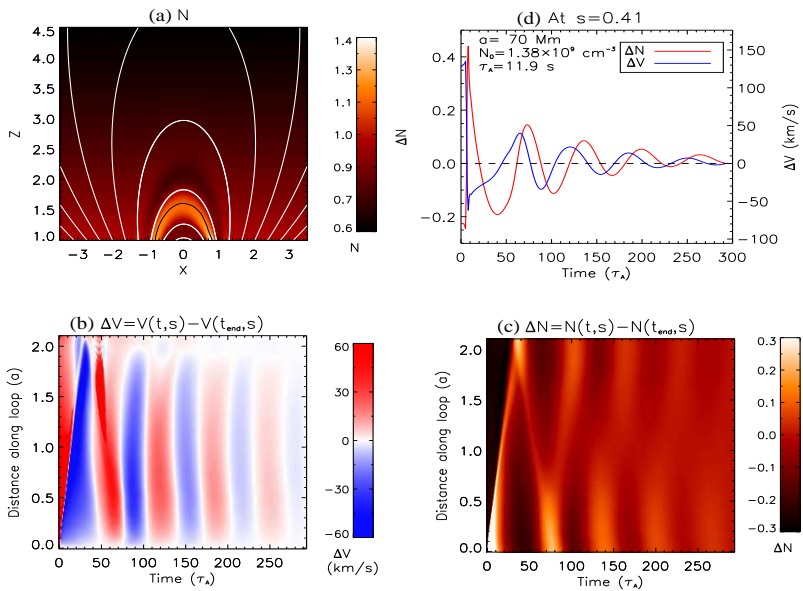}
         \caption{Isothermal 3D MHD simulations of slow-mode waves excited by the impulsive onset of steady flows (supplemental material of \citealt{ofm12}). Panels from top-left in counter-clockwise direction: (a) Snapshot of the density in the $xz$-plane (at $y$=0) at $t=280\tau_A$ with steady inflow ($V_0=0.05V_A$), injected at the right footpoint. Superposed are the magnetic field lines (white) and a cut along the loop (black). (b) Time distance plot of the perturbed velocity ($\Delta {V}=V(s, t)-V(s, t_{end})$) for the cut along the loop (marked in (a)), where $V(s, t)$ and $V(s, t_{end})$ are the spatial profiles of the total velocity in the $xz$-plane at a time $t$ and $t_{end}=294\tau_A$, respectively, and $V(s, t)$ takes the sign of $V_x$. (c) Same as (b) but for the perturbed density. (d) Time profiles of the perturbed density and velocity at the loop position $s$=0.41 from the right footpoint, where $s$ in units of $a$=70 Mm. A quarter-period phase shift set up between $\Delta{V}$ and $\Delta{N}$ at $t\approx{60}\tau_A$ indicates the formation of the standing mode.  }
         \label{fig:m3d}
 \end{figure}

In particular, \citet{ofm12} studied the excitation of waves by injected flows at the lower coronal boundary in realistic AR structures using 3D MHD modeling. The model AR was constructed by using a dipole (potential) magnetic field together with gravitationally stratified density and steady or periodic upflows in various locations of the magnetic loops' footpoints. The model was an extension of 3D resistive and isothermal MHD model of a bipolar AR with gravitationally stratified density initially developed by \citet{ofmt02} to study waves in ARs, and since then used in many studies (see, recently, \citealt{ofm18} and references within). \citet{ofm12} found that the upflows at the boundaries produce siphon flows and higher density loops in the model AR. The impulsive flow injection leads to oscillations and excitation of coupled MHD waves, in the form of fast and slow magnetoacoustic waves. In particular, the impulsive injection of (subsequently) steady flows produces slow magnetoacoustic waves that travel along the loops. They found that the slow-mode waves quickly transform to standing oscillations.

\begin{figure}
         \centering
         \includegraphics[width=0.9\textwidth]{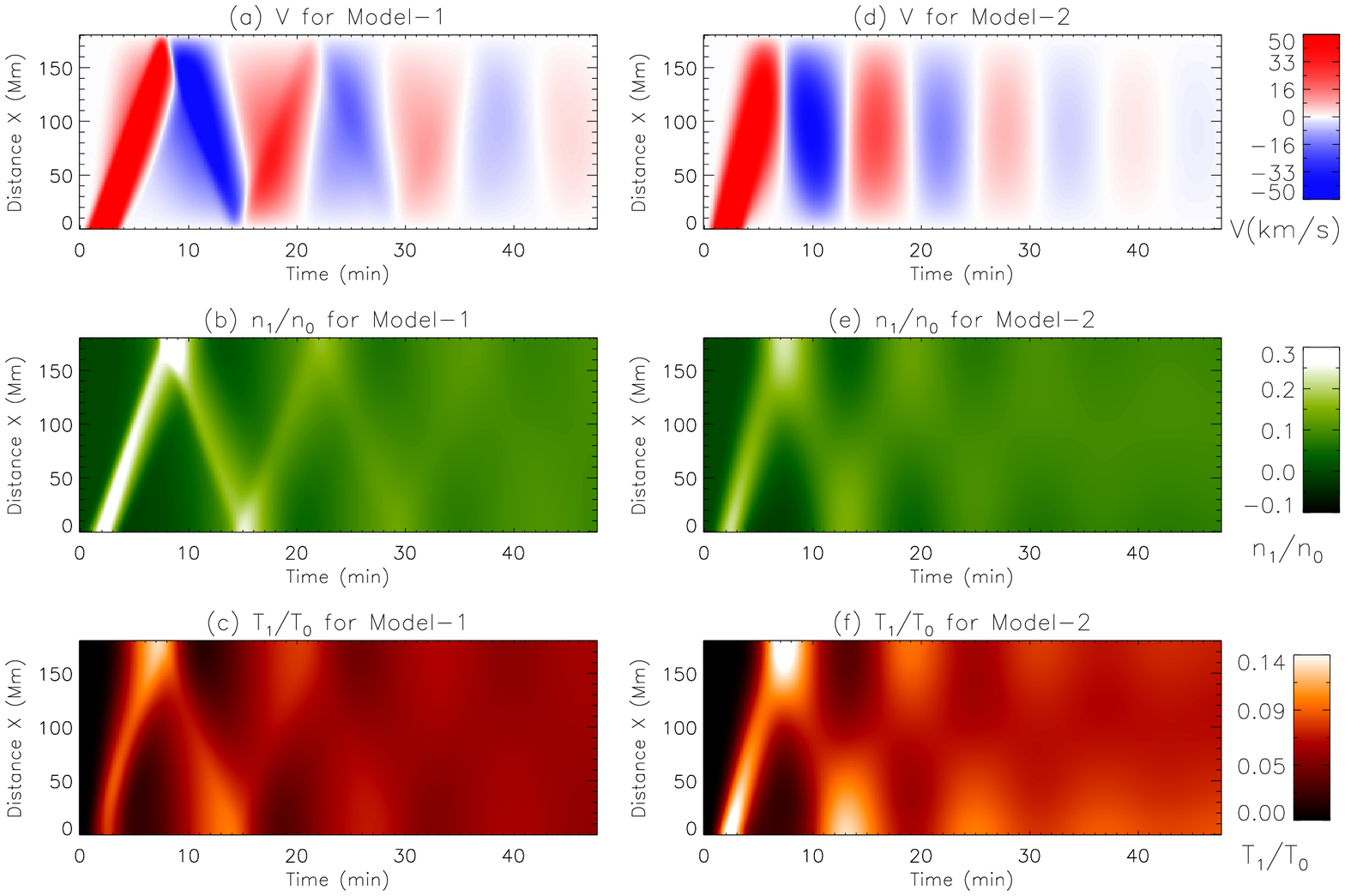}
         \caption{Comparison between simulations of the slow-mode wave excitation by a flow pulse at one footpoint using a 1D MHD loop model in the two cases \citep[from][]{wan18}. Left panels: Model 1 with the classical thermal conduction and viscosity coefficients. Right panel: Model 2 with the observation-determined transport coefficients, i.e., no thermal conduction and 15 times enhanced viscosity. (a)-(c) Time-distance maps of velocity, perturbed density, and perturbed temperature for Model 1. (d)-(f) Same as (a)-(c) but for Model 2. }
         \label{fig2_ofman}
 \end{figure}

Figure~\ref{fig:m3d} shows the results of 3D MHD modeling from \citet{ofm12} for the case with the isothermal background $T_0=6.3$ MK, stratified density with $n_0=1.38\times10^9$ cm$^{-3}$ at $z$=1 in units of $a$=70 Mm and steady inflow with velocity magnitude $V_0=0.05V_A$ in normalized units. In Fig.~\ref{fig:m3d}a the density of the loop formed by the upflows is shown in an $xz$ plane cut at the center of the AR ($y=0$) at time $t=280\tau_A$. The effects of the initially impulsive and subsequently steady flow are evident in the formation of the higher density loops and in the excitation of the oscillation (Fig.~\ref{fig:m3d}b-d). In Fig.~\ref{fig:m3d}b the velocity perturbations along the loop as function of time are shown in the cut marked by the black line in Fig.~\ref{fig:m3d}a. The density perturbations along the same cut are shown in Fig.~\ref{fig:m3d}c, and the time dependent oscillations at a point inside the loop are shown in Fig.~\ref{fig:m3d}d. The spatio-temporal patterns of the perturbed velocity and density indicate that a standing slow mode is set up within about one wave period. The formation of the standing slow magnetoacoustic wave is also evident from the time dependences of the velocity and density perturbations that become quarter-period phase shifted (Fig.~\ref{fig:m3d}d), in agreement with the theoretical prediction (see Sect.~\ref{sst:fm}). Based on their 3D MHD modeling results,  \citet{ofm12} concluded that impulsive events (such as flares) that result in upflows can explain the origin of the observed slow-mode waves in AR loops. The quick formation of the standing wave could be related to transverse structuring and wave leakage in the curved geometry \citep[e.g.][]{ogr09} in addition to the damping by thermal conduction and viscosity \citep{wan19}.

The obvious advantage of the 1D model vs. 3D model is the much smaller computational requirements for the same numerical parameters, which facilitate the use of realistic dissipation coefficients in parametric studies.  Recently, \citet{wan18} and  \citet{wan19} used 1D nonlinear, viscous, thermally conductive MHD model to study the excitation and damping of slow magnetoacoustic waves in a flaring hot loop observed on 2013 December 28 with SDO/AIA. By applying the coronal seismology technique to this event (see Sect.~\ref{sst:tc}), they determined the transport coefficients in hot loop plasma at $\sim$10 MK and revealed strong suppression of thermal conduction with significant enhancement of compressive viscosity by more than an order of magnitude. Using parametric study of the dissipation coefficients with the 1D MHD model, \citet{wan19} found that the thermal conduction was suppressed by a factor of 3 compared to the classical value and the compressive viscosity was enhanced by a factor of 10 in the observed loop.

\begin{figure}
         \centering
         \includegraphics[width=0.9\textwidth]{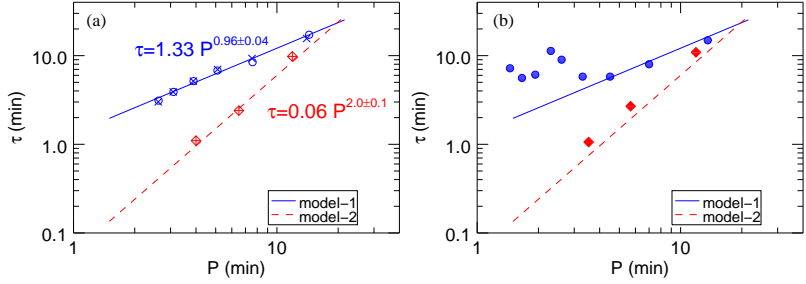}
         \caption{Dependence of the damping time on the period of the velocity component of the slow-mode waves for Models 1 and 2 \citep[from][]{wan18}. (a) Model 1 (circles and crosses) and Model 2 (diamonds and pluses). The circles and diamonds are for cases with initial velocity amplitude $V_0=0.23 C_s$, while $\times$ and $+$ are for case with $V_0=0.023 C_s$. The best-fit power laws are shown. (b) The damping time vs. period for various Fourier components of the waves for Model 1 (filled circles) and Model 2 (filled diamonds) with best-fit power laws.}
         \label{fig3_ofman}
 \end{figure}

Another striking result is that they found that the model with the seismology-determined transport coefficients can self-consistently produce the standing slow-mode wave as quickly (within one period) as observed (Model 2; Figs.~\ref{fig2_ofman}d-f), whereas the model with the classical transport coefficients produces initially propagating slow-mode waves that need many reflections to form a  standing wave (Model 1; Figs.~\ref{fig2_ofman}a-c). Using the 1D MHD simulations \citet{wan18} analyzed the frequency dependence of harmonic waves dissipation and demonstrated the underlying cause for the difference of the two models. For Model 2 the scaling law relation between damping time and wave period is close to $\tau\propto{P^2}$, while Model 1 produces $\tau\propto{P}$ (see Fig.~\ref{fig3_ofman} and a discussion in Sect.~\ref{sss:nme}). Such relations suggest that the anomalous viscosity enhancement facilitates the dissipation of higher harmonic components in the initial perturbation pulse, so that the the fundamental standing mode could quickly form. This dependence on the dissipation coefficients may provide an explanation for the excitation of both standing and reflective longitudinal oscillations observed with SDO/AIA in different events with different loop conditions. When the viscosity dominates in wave damping (corresponding to Model 2), the fundamental standing wave is preferentially excited, whereas when the thermal conduction is the dominant damping mechanism (Model 1), the reflected propagating waves are excited. Thus, the viscous, thermally conductive MHD model provides a new coronal seismology method for the determination of thermal conduction and compressive viscosity in a hot coronal loop (see Sect.~\ref{sst:tc}). 

\subsection{Modeling of reflecting waves}
\label{sst:mrw}

\begin{figure}
\centering
\includegraphics[width=0.9\textwidth]{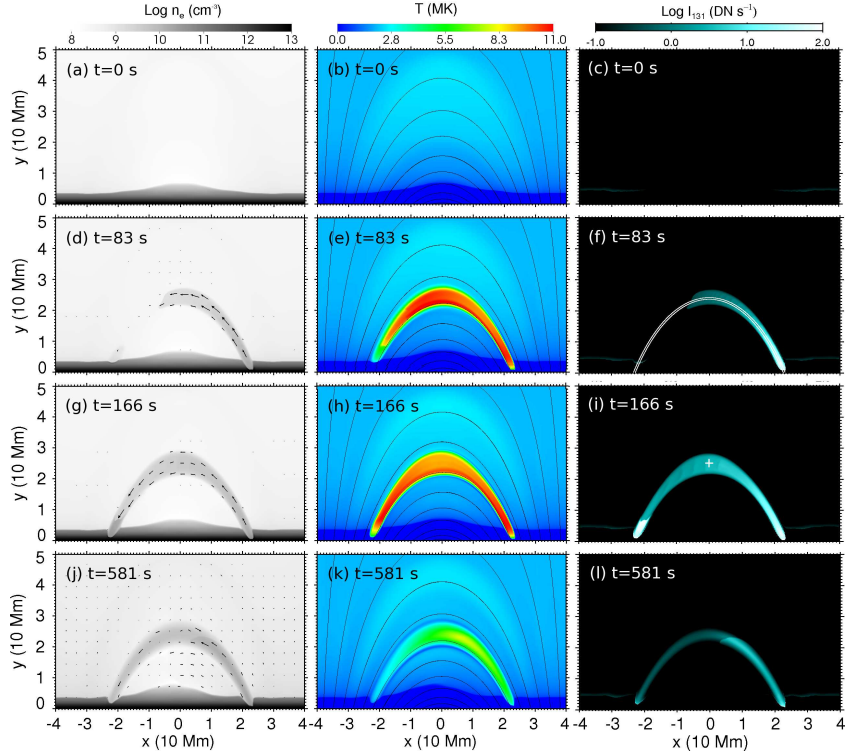}
\caption{Temporal evolution of number density (left column), temperature (middle column), and synthesized AIA 131 \AA\ emission (right column) images at $t =0$, 83, 166 and 581 s, respectively \citep[from][]{fang15}.}
 \label{fig:fang2015_fig1}
\end{figure}

\citet{fang15} used a 2.5D MHD code with radiative cooling and thermal conduction,  and simulated the excitation and propagation of propagating slow-mode waves in a closed coronal loop. An arcade of magnetic field was set up to mimic a bipolar linear force-free field (Fig.~\ref{fig:fang2015_fig1}b). A stratified atmosphere was added into this initial magnetic configuration, with a model for the chromosphere, transition region, and corona. An impulsive heating was applied at one footpoint of a thin flux  tube, the plasma was heated to a high coronal temperature and filled the flux tube rapidly (Figs.~\ref{fig:fang2015_fig1}d-i).  Here we note that the thermal front propagates faster than the density front, \citet{wan18} reported observational evidence that the loop is heated by a thermal front that precedes the propagating waves.

\begin{figure}
\centering
\includegraphics[width=0.9\textwidth,height=0.3\textheight]{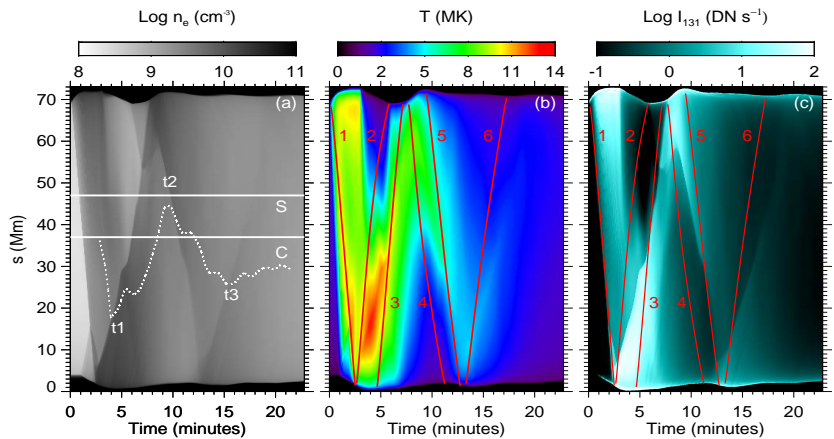}
\caption{Variation of electron number density, temperature, and the synthesized AIA 131 \AA\ emission along the loop \citep[from][]{fang15}. The six red solid lines in panels (b) and (c) show the path of acoustic waves propagating at the local sound speed. }
 \label{fig:fang2015_fig2}
\end{figure}

To study the properties of propagating slow-mode waves and flows, \citet{fang15} traced the evolution of plasma density, temperature, and synthesized AIA 131 \AA{} emission intensity along the loop (Fig.~\ref{fig:fang2015_fig2}). One could see that near the footpoints, the theoretical trajectory of the sound waves deviates significantly from the phase speed of the simulated wave fronts in plasma density, temperature, and AIA 131 \AA{} emission intensity, this means in this region, the propagating front is a mixture of plasma flow and slow-mode wave. Whereas in contrast, close to the loop apex, this propagating front is almost in parallel with the trajectory of the slow-mode waves, this means the wave component dominates in this region. The simulation of \citet{fang15} suggests that one has to be cautious in interpreting the propagating reflected wave patterns observed in coronal loops, because the properties of the slow-mode waves (e.g., propagating speed and amplitude) could be affected by the background mass flow. This should be significant when the flow speed is close to sound speed. This also applies to the measurement of wave damping. We should also note that compressive viscosity is not included in the MHD model of \citet{fang15}. The viscosity may play an important role in reducing the effect of flow on the waves and also suppressing the nonlinear effect which is obviously seen in their simulations, this may lead to the significant distinction between the synthetic and observed intensity oscillations \citep[see][for some discussions]{wan18}.

\section{Damping mechanisms}
\label{sst:dm}

Both standing and propagating slow-mode waves observed in coronal loops exhibit fast damping. The issue of damping has attracted a remarkable attention since the discovery of these waves. Table~\ref{tab:dmp} lists most of the relevant studies in the literature that are dedicated to this problem. Many damping mechanisms were investigated, including non-adiabatic effects such as thermal conduction, compressive viscosity, and radiation (see Sect.~\ref{sss:nme}), the nonlinearity, the cooling loop background, the wave-caused heating/cooling imbalance (see Sect.~\ref{sss:whc}), plasma non-uniformities, and other effects such as loop geometry \citep[e.g., loop expansion and curvature; see][]{dem04a,ogr09}, wave leakage \citep[e.g., in footpoints and the corona; see][]{selw07,ogr09}, and magnetic effects \citep[e.g., mode coupling and obliqueness;][]{dem04b,afa15}. Various methods were used to analyze and compare the importance of different mechanisms under various coronal conditions. The basic and also the most commonly used method is to derive the dispersion relation from linearized MHD equations for a uniform loop model including a single or multiple dissipation mechanisms such as thermal conduction, viscosity, and radiative cooling \citep[e.g.][]{dem03, dem04a, pan06, sig07}. Analytic or numerical solutions of the dispersion relation allow us to readily examine the dependence of wave frequency and damping rate on the physical parameters of plasma such as the equilibrium temperature ($T_0$) and density ($n_0$) in wide ranges. The main limitation of this approach is that it neglects the effects of wave nonlinearity, which can affect the wave propagation and damping \citep[e.g.,][]{ofm00,ofm02,wan19}. By applying WKB theory \citep[see][]{bend78} to a time-dependent equilibrium (e.g., assuming cooling of the background plasma due to thermal conduction and optically thin radiation), a time-dependent dispersion relation and analytic solutions for the time-dependent amplitude of waves can be obtained \citep[e.g.][]{mort10,erd11,alg13}. When assuming that the nonlinearity, dissipation, and reflection effects are weak, the WKB theory can be used to derive a generalized Burgers equation that governs the evolution of the oscillations in a propagating mode \citep[e.g.][]{nak00,afa15} or a standing mode \citep[e.g.][]{rud13,kums16}. Linear and nonlinear MHD simulations are often used to study the wave damping in more realistic solar conditions, by including various effects such as multiple dissipation mechanisms, magnetic field geometries, gravitational stratification, transverse and longitudinal inhomogeneous plasma structuring, nonlinear mode coupling, wave leakage, and so on (see the references in Table~\ref{tab:dmp}). This allows direct comparison of modeling with observations.

The observed propagating and standing slow-mode waves in coronal loops are often studied using similar theoretical approaches  but based on different models. This is because they are dissipated essentially by same physical processes, however, in distinctly different physical conditions and magnetic geometry structures. The EUV propagating waves are observed in the footpoints of large, warm (1$-$2 MK) fan loops, with a continuous quasi-periodic or broadband driver and small relative amplitudes of typically 3$-$4\% of the background intensity \citep[e.g.][]{dem02,mcew06,wan09}, while the SUMER standing waves are observed in hot ($>$6 MK) flaring loops, which are impulsively generated by a single flow pulse with large velocity amplitudes on average about 10$-$20\% of the loop sound speed \citep{wan03a,wan05,wan07}. Loop expansion appears to play an important role in damping the propagating waves \citep{dem04a,mars11}, while the loop length and curvature may be important for damping of the standing waves \citep{ogr09}. In the present review we focus primarily on damping of observed standing waves in the hot coronal loops. 

\begin{table*}
\begin{threeparttable}
\caption{List of studies on damping mechanisms of slow-mode waves in coronal loops}
\label{tab:dmp}       
\begin{tabular}{lllll}
\hline\noalign{\smallskip}
Method & Mechanism\tnote{a} & Non-uniform\tnote{b} & Mode\tnote{c} & References  \\
\noalign{\smallskip}\hline\noalign{\smallskip}
Linear & $\kappa_{\|}$ & U & P, S & \citet{dem03,owe09}\\
Theory & & & & \citet{krish14}\\
       & $\eta$ & U & S & \citet{sig07}\\
       & $Q$ & U & P, S & \citet{dem04a,sig07}\\
       & Ideal & $A(s)$+$G(s)$ & P & \citet{dem04a}\\
       & $\eta$ & $\rho_0(x)$, $B_0(x)$ & P & \citet{dem04b} \\
       & $\eta$ & $G(s)$ & $S$ & \citet{sig07}\\
       & $\kappa_{\|}$+$\eta$+$Q$ & U & S & \citet{pan06} \\
       & $\kappa_{\|}$+$\eta$+$Q$ & $G(s)$+$T_0(s)$+$\rho_0(s)$ & S & \citet{abed12}\\
       & $T_0(t)$+$Q$ & U & P & \citet{mort10} \\
       & $T_0(t)$+$\kappa_{\|}$ & $T_0(s)$+$\rho_0(s)$ & P & \citet{erd11}\\
       & $T_0(t)$+$\kappa_{\|}$ & U & P, S & \citet{alg13,alg14}\\
       & $T_0(t)$+$\kappa_{\|}$+$Q$ & U & S & \citet{alg15}\\
       & $T_0(t)$+$\kappa_{\|}$+$\eta$ & U & S & \citet{bah18} \\
       & $WQ$+$\kappa_{\|}$+$\eta$ & U & S & \citet{kums16} \\
       & $WQ$+$\kappa_{\|}$ & U & S & \citet{kolot19} \\
       & $V_0$+$\eta$ & U & P & \citet{kumn16} \\
\noalign{\smallskip}\hline\noalign{\smallskip}
Nonlinear & $\kappa_{\|}$+$\eta$ & $G(s)$ & P & \citet{nak00}\\
Theory & $\kappa_{\|}$+$\eta$ & U & S & \citet{rud13}\\
       & $\eta$+$B$ & U & $P$ & \citet{afa15}\\ 
       & $WQ$+$\kappa_{\|}$+$\eta$ & U & S & \citet{kums16}\\
       & $WQ$+$\kappa_{\|}$+$\eta$+$B$ & U & P & \citet{nak17}\\
\noalign{\smallskip}\hline\noalign{\smallskip}
Numerical & $\kappa_{\|}$+$\eta$ & U & P, S & \citet{dem03}\\
Simulation & $\kappa_{\|}$+$Q$ & U & P & \citet{dem04a}\\
(linear)   & $\kappa_{\|}$ & $G(s)$+$A(s)$ & P & \citet{dem04a}\\
          & $\eta$ & $B_0(x)$, $\rho_0(x)$ & P & \citet{dem04b}\\
          & $\kappa_{\|}$+$\eta$ & $G(s)$ & S & \citet{sig07}\\
	& $\kappa_{\|}$+$\eta$+$Q$ & $G(s)$ & P & \citet{sig09}\\	
     	& $V_0$+$\eta$ & U & S & \citet{kumn16} \\
\noalign{\smallskip}\hline\noalign{\smallskip}
Numerical & $\kappa_{\|}$+$\eta$ & U & S & \citet{ofm02},\citet{wan18}\\
Simulation & & & & \citet{wan19} \\
(nonlinear) & $\kappa_{\|}$+$Q$ & $G(s)$+$T_0(s)$+$A(s)$ & $P$ & \citet{klim04}\\
 & $\kappa_{\|}$+$\eta$ & $G(s)$ & S & \citet{mend04,sig07}\\
     & $\kappa_{\|}$+$\eta$+$Q$ & $G(s)$+$T_0(s)$ & S & \citet{erd08}\\
     & $\kappa_{\|}$+$Q$+leakage & $\rho_0(s)$+$T_0(s)$ & P, S & \citet{selw05, jeli09}\\
     & $\kappa_{\|}$+Shock & U & S & \citet{verw08}\\
     & $Q$/$Q_{N}$+$\kappa_{\|}$ & $G(s)$+$\rho_0(s)$+$T_0(s)$ & S & \citet{brad08}\\
     & $\kappa_{\|}$ & $G(s)$+$\rho_0(s)$+$T_0(s)$ & P &  \citet{owe09}\\
     & Ideal+leakage & $\rho_0(x,z)$+$B_0(x,z)$ & P, S &  \citet{selw07, ogr09}\\
     & $Q$ & $G(z)$+$\rho_0(z)$ & P & \citet{prov18}\\ 
     & & +$T_0(z)$+$B(x,y,z)$ & & \\ 
\noalign{\smallskip}\hline
\end{tabular}
  \begin{tablenotes}
  \item[a] $\kappa_{\|}$ stands for thermal conduction, $\eta$ for compressive viscosity, $Q$ ($Q_{N}$) for optically-thin radiation in equilibrium (non-equilibrium) ionization balance, $T_0(t)$ for cooling background, $WQ$ for wave-caused heating/cooling imbalance, $B$ for obliqueness and magnetic effects, and $V_0$ for the steady flow.
  \item[b] U represents the loop model with uniform equilibrium, $G(s)$ (or $G(z)$) for gravitational stratification, $A(s)$ for loop expansion, $\rho_0(s)$ and $T_0(s)$ for non-uniform equilibrium density and temperature along the loop, $\rho_0(x)$ and $B_0(x)$ for non-uniform density and magnetic field in the transverse direction, $\rho_0(x,z)$ and $B_0(x,z)$ for non-uniform 2D distributions, and $B(x,y,z)$ for non-uniform 3D distribution.
  \item[c] P stands for propagating wave, S for standing wave. 
  \end{tablenotes}
\end{threeparttable}
\end{table*}

\subsection{Non-ideal MHD effects}
\label{sss:nme}
Thermal conduction, compressive viscosity, and optically thin radiation are the most nominated mechanism for damping of slow-mode waves, and have been studied intensively. However, despite the investment of much effort the interpretations on fast damping of standing slow-mode waves in hot coronal loops still do not reach a concord. Based on a 1D nonlinear MHD modeling guided by SUMER observations, \citet{ofm02} first suggested that thermal conduction is the dominant damping mechanism of standing slow magnetoacoustic waves. They found that the damping rate due to compressive viscosity alone is too weak to account for the observed decay times. However, some later studies based on linear analytical and numerical simulations have shown that thermal conduction alone results in the density and velocity waves with slower damping, insufficient to explain some observations, and that the viscous dissipation is required to be added to reproduce the rapidly damping as observed, particularly in shorter and hotter loops \citep{mend04,sig07,abed12}. By studying the evolution of oscillations in a slowly cooling coronal loop using the WKB method, \citet{bah18} also concluded that in hot loops the efficiency of compressive viscosity in damping slow-mode waves is comparable to that of thermal conduction. \citet{pan06} examined the effect of radiation on wave damping from solutions of the dispersion relation derived in the presence of thermal conduction, viscosity, and optically thin radiation, and found that for strong-damped oscillations ($\tau/P\sim{1}$) in a lower density condition ($n_0=10^8-10^9$ cm$^{-3}$), the radiative effect is negligible compared to that of thermal conduction and viscosity, whereas for weak-damped oscillations ($\tau/P\geq{1}$) at higher density ($n_0\geq{5\times 10^9}$ cm$^{-3}$), the additional dissipation due to radiation becomes evident. The conclusion that radiative cooling is an insignificant mechanism for dissipation of slow-mode waves in typical hot coronal loops was also drawn by some other theoretical and numerical studies based on 1D HD model \citep{sig07,abed12} and 3D MHD model \citep{prov18}. In addition, \citet{brad08} inspected the influence of non-equilibrium ionization balance on the importance of optically thin radiation in damping, and found that this effect is generally weak for hot loops (e.g., reducing damping times by less than 5\% at $T_0$=8 MK compared to the equilibrium case).

By revisiting the dispersion relations for the slow-mode wave dissipation due to thermal conduction, viscosity, and radiation in a uniform loop model in this section, we show that the efficiencies of these three mechanisms are sensitive to the choice of loop physical parameters ($n_0$, $T_0$, and $L$). This suggests that some inconsistent conclusions in the studies mentioned above likely lie in discrepancies of the physical parameters used in their models.

 \begin{figure}
         \centering
         \includegraphics[width=1.0\textwidth]{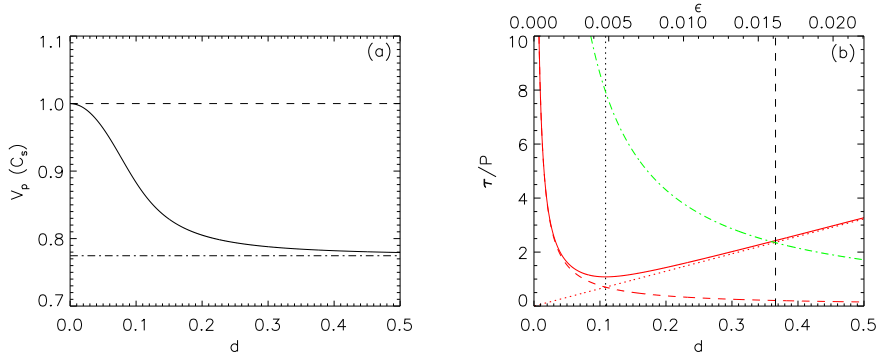}     
         \caption{(a) Phase speed of slow-mode waves (in units of the adiabatic sound speed $C_s$) as a function of thermal ratio $d$. The dashed line indicates $C_s$, and the dotted dashed line indicates the isothermal sound speed $C_0=\gamma^{-1/2}C_s$. (b) Ratio of damping time to wave period $\tau/P$ as a function of thermal ratio $d$ (solid line). The dashed line represents its approximation when $d\ll{0.1}$ while the dotted line the case when $d\gg{0.1}$. The variation of $\tau/P$ against viscous ratio $\epsilon$ is plotted in the dot-dashed line. The vertical dotted line indicates the minimum point of $\tau/P$ for thermal conduction, and the vertical dashed line indicates the crossing point of the curves for thermal conduction and viscosity.}
         \label{fig:tpd}
 \end{figure}

\noindent
(1) Thermal conduction

The importance of thermal conduction in wave dissipation can be quantified by the thermal ratio $d$ as defined in \citet{dem03},
\begin{equation}
 d=\frac{1}{\gamma}\frac{P_0}{\tau_{\rm cond}}=\frac{(\gamma-1)\kappa_\|T_0 \rho_0}{{\gamma}^2p_0^2 P_0}=\frac{\mu m_p (\gamma-1)\kappa_0}{2\gamma^2 k_B^2}\left(\frac{T_0^{3/2}}{n_0 P_0}\right),
 \label{equ:d}
\end{equation}
where $P_0=\lambda/C_s$ and $\tau_{\rm cond}$ is the thermal conduction timescale, $\kappa_\|=\kappa_0T_0^{5/2}$ is the classical Spitzer thermal conductivity parallel to the magnetic field (with $\kappa_0=7.8\times10^{-7} {\rm erg~cm}^{-1} {\rm s}^{-1} {\rm K}^{-7/2}$), $m_p$ the proton mass, and $\mu$=0.6. For a fundamental mode, the wavelength $\lambda=2L$, so the dependence of $d$ on the loop physical parameters (in cgs units) can be written as,
\begin{equation}
  d=4.93\left(\frac{T_0^{3/2}}{n_0P_0}\right)=3.75\times{10^4}\left(\frac{T_0^2}{n_0L}\right).
 \label{equ:dcgs}
\end{equation}
For thermal conduction as the only damping mechanism, a dispersion relation can be derived from the linearized MHD equations under the assumption of all disturbances in the form $e^{i(\omega{t}-kz)}$ \citep[e.g.][]{dem03,krish14},
\begin{equation}
\omega^3-(i\,\gamma\,dP_0C_s^2k^2)\omega^2-(C_s^2k^2)\omega+ i\,dP_0C_s^4k^4= 0.
\label{equ:cdsp}
\end{equation}
Because of $C_s k=2\pi/P_0$, the equation with a fixed timescale $P_0$ for the wave frequency $\omega=\omega_r+i\omega_i$ can be solved numerically \citep{wan19}. Figure~\ref{fig:tpd} shows our calculated results for the phase speed $V_p=\omega_r/k$ and the ratio of damping time to wave period $\tau/P=\omega_r/2\pi\omega_i$ as a function of thermal ratio $d$. As found in \citet{dem03}, the slow-mode waves have a minimum in $\tau/P$ and $\tau$ due to thermal conductivity. The calculations find $(\tau/P)_{\rm min}\approx$1.1 at $d_m\approx$0.11 and $\tau_{\rm min}/P_0\approx$1.2 at $d_m\approx$0.10, which are independent of the choice of $P_0$. The slow-mode waves propagate at a near-adiabatic sound speed when $d\ll 0.1$, while at a near-isothermal sound speed when $d\gg 0.1$ (see Fig.~\ref{fig:tpd}a). \citet{krish14} showed that the dispersion relation can be approximated to a simple form in the weak or strong conduction regime. 

(i) In the weak thermal conduction ($d\ll 0.1$) approximation
\begin{equation}
 \omega=kC_s+ i\left(\frac{\gamma-1}{2}\right)dP_0k^2C_s^2,
\label{equ:wtc}
\end{equation}
then we have
\begin{eqnarray}
V_p &=& \frac{\omega_r}{k}=C_s, \\
P &=& \frac{2\pi}{\omega_r}=P_0, \\
\tau &=& \frac{1}{\omega_i}=\frac{1}{2\pi^2(\gamma-1)}\left(\frac{P_0}{d}\right)\propto \frac{n_0P_0^2}{T_0^{3/2}} \propto P^2, \label{equ:wtt} \\
\frac{\tau}{P} &=& \frac{1}{2\pi^2(\gamma-1)}\left(\frac{1}{d}\right) \propto \frac{n_0L}{T_0^2}.
\label{equ:wtp}
\end{eqnarray}

(ii) In the strong thermal conduction ($d\gg 0.1$) approximation
\begin{equation}
 \omega=\gamma^{-1/2}kC_s+ i\frac{\gamma-1}{2\gamma^2 dP_0},
\label{equ:stc}
\end{equation}
then we have
\begin{eqnarray}
V_p &=& \frac{\omega_r}{k}=\frac{C_s}{\gamma^{1/2}} \equiv C_0, \\
P &=& \frac{2\pi}{\omega_r}=\gamma^{1/2}P_0, \\
\tau &=& \frac{1}{\omega_i}=\frac{2\gamma^2dP_0}{\gamma-1}  \propto \frac{T_0^{3/2}}{n_0} \propto P^0, \label{equ:stt} \\
\frac{\tau}{P} &=& \left(\frac{2\gamma^{3/2}}{\gamma-1}\right)d \propto \frac{T_0^2}{n_0L}.
\label{equ:stp}
\end{eqnarray}
Equations~(\ref{equ:wtp}) and~(\ref{equ:stp}) are the two asymptotic solutions to $\tau/P$ (see Fig.~\ref{fig:tpd}b), from their crossing point we estimate $d_m=1/(2\pi\gamma^{3/4})\approx 0.11$, where $\tau/P$ reaches the minimum. Equation~(\ref{equ:wtt}) indicates $\tau\propto P^2$ when $d\ll 0.1$ while Equation~(\ref{equ:stt}) indicates $\tau\propto P^0$ when $d\gg 0.1$, in agreement with the result in \citet{krish14}. \\

\begin{table*}
\begin{threeparttable}
\caption{Parameters for thermal ratio ($d$), viscous ratio ($\epsilon$), and radiation ratio ($r$) as functions of the equilibrium density ($n_0$), temperature ($T_0$), or loop length ($L$) for the fundamental standing slow-mode waves\tnote{a}}
     \label{tab:der}       
\begin{tabular}{llll}
\hline\noalign{\smallskip}
Parameters & $d$ & $\epsilon$ & $r$ \\
\noalign{\smallskip}\hline\noalign{\smallskip}
$n_0=10^8\rightarrow{10^{11}}$ (cm$^{-3}$) & $1.4\rightarrow 1.4\times{10^{-3}}$ & $0.074 \rightarrow 7.4\times{10^{-5}}$ & $5.5\times{10^{-4}}\rightarrow{0.55}$ \\
$T_0=2\rightarrow{20}$ (MK) & $2.7\times{10^{-3}}\rightarrow 0.27$ & $1.4\times{10^{-4}}\rightarrow 0.014$ & $0.37\rightarrow 2.5\times{10^{-3}}$\\
$L=20\rightarrow{400}$ (Mm) & $0.49\rightarrow 0.024$ & $0.026\rightarrow 1.3\times{10^{-3}}$ & $1.6\times{10^{-3}}\rightarrow 0.032$\\
\noalign{\smallskip}\hline
\end{tabular}
\begin{tablenotes}
\item[a] The physical parameters $n_0=2.6\times{10^9}$ cm$^{-3}$, $T_0$=9 MK, and $L$=180 Mm are taken from measurements of an AIA wave event studied in \citet{wan15}, which yield $d$=0.065, $\epsilon$=0.0029, and $r$=0.014. The listed values in the table for $d$, $\epsilon$, and $r$ are calculated for a loop with $n_0$, $T_0$, and $L$ by varying one of the parameters.
\end{tablenotes}
\end{threeparttable}
\end{table*}

\noindent
(2) Compressive viscosity

The dispersion relation for dissipation of slow-mode waves by viscosity alone can be obtained as \citep[e.g.][]{ofm00,sig07,wan19},
\begin{equation}
\omega^2-i\left(\frac{4}{3}\epsilon P_0 C_s^2 k^2\right)\omega-C_s^2 k^2=0, 
\label{equ:vdpa}
\end{equation}
with the solutions
\begin{equation}
\omega=\pm\,{kC_s}\left(1-\frac{4}{9}\epsilon^2 P_0^2 C_s^2 k^2\right)^{1/2}+i\left(\frac{2}{3}\epsilon P_0 C_s^2 k^2\right).
\label{equ:vdpb}
\end{equation}
Here $\epsilon$ is the viscous ratio, defined as,
\begin{equation}
 \epsilon=\frac{1}{R}=\frac{\eta_0}{\rho_0 C_s^2 P_0},
 \label{equ:vis}
\end{equation}
where $R$ is the Reynolds number, $\eta_0=\bar{\eta}T_0^{5/2}$ is the classical Braginskii compressive viscosity coefficient (with $\bar{\eta}=10^{-16}{\rm g~cm}^{-1}\,{\rm s}^{-1}{\rm K}^{-5/2}$), and $P_0=\lambda/C_s$. For the fundamental mode, $\lambda=2L$, then the viscous ratio $\epsilon$ can be expressed in the form of $T_0$, $n_0$, and $P_0$ (or $L$) in cgs units as
\begin{equation}
 \epsilon=0.217\left(\frac{T_0^{3/2}}{n_0P_0}\right)=1.65\times{10^3}\left(\frac{T_0^2}{n_0L}\right).
\label{equ:vcgs}
\end{equation}
We then have
\begin{eqnarray}
 P &=& \frac{2\pi}{\omega_r}= P_0\left(1-\frac{16\pi^2\epsilon^2}{9}\right)^{-1/2}\approx{P_0},  \label{equ:pv}\\
 \tau &=& \frac{1}{\omega_i}=\frac{3}{8\pi^2}\left(\frac{P_0}{\epsilon}\right) \propto \frac{n_0P_0^2}{T_0^{3/2}} \propto{P^2},    \label{equ:vtt} \\
 \frac{\tau}{P} &=& \frac{3}{8\pi^2}\left(\frac{1}{\epsilon}\right) \propto\frac{n_0L}{T_0^2}.
 \label{equ:tpv}
\end{eqnarray}
Equation~(\ref{equ:pv}) suggests that the effect of viscosity on the wave period is negligible (of the second-order of smallness), because the viscous ratio is small ($\epsilon\lesssim{10^{-2}}$) in the hot coronal condition (see Table~\ref{tab:der}). Since the dependence of viscous and thermal ratios on the parameters ($T_0$, $n_0$, and $P_0$ (or $L$)) follows the same form (comparing Eq.~\ref{equ:vcgs} with Eq.~\ref{equ:dcgs}; see also \citealt{mac10}), this implies that the ratio $d/\epsilon$ is a constant ($\approx$22.7). It allows to compare the variation of $\tau/P$ against $d$ for thermal conduction damping with that of $\tau/P$ against $\epsilon$ for viscous damping in the same plot (see Fig.~\ref{fig:tpd}b). At $d_m$=0.11, we obtain $\epsilon=d_m/22.7=4.8\times 10^{-3}$ and $(\tau/P)_{\rm visc}$=7.9 using Eq.~(\ref{equ:tpv}). In the case of $d\ll{0.1}$, we can obtain using Eqs.~(\ref{equ:wtp}) and ~(\ref{equ:tpv}) that
\begin{equation}
\frac{(\tau/P)_{\rm visc}}{(\tau/P)_{\rm cond}}=\frac{\tau_{\rm visc}}{\tau_{\rm cond}}=\frac{3(\gamma-1)}{4}\left(\frac{d}{\epsilon}\right)\approx 11.  
  \label{equ:ctv}
\end{equation}
These estimates indicate that $d\lesssim$0.1 is a sufficient condition for thermal conduction dominating over viscosity in wave dissipation. In the case of $d>{0.1}$ we estimate $d_c\approx$0.37 and $(\tau/P)_c\approx$2.4 for the crossing point between the two curves for thermal conduction and viscosity using Eqs.~(\ref{equ:stp}) and ~(\ref{equ:tpv}). This suggests that the viscous damping begins to dominate over the thermal conduction damping when $d>0.37$ (or $\epsilon>$0.016; e.g. in higher harmonics or shorter hot loops).

\begin{figure}
        \centering
        \includegraphics[width=1.0\textwidth]{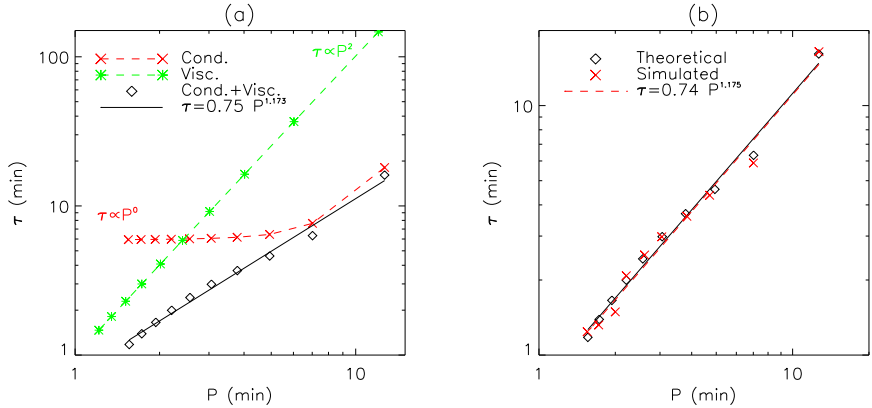}
        \caption{Damping of the slow-mode wave by thermal conduction and compressive viscosity in a hot loop with $T_0$=9 MK, $n_0=2.6\times 10^9$ cm$^{-3}$, and $L$=180 Mm. (a) Variation of the damping time as a function of the wave period for the harmonics $n=1-10$. The case for the wave dissipation by thermal conduction (viscosity) alone is plotted with the crosses (asterisks). The solid line is the best-fit scaling for the case with the combined damping effects (diamonds). (b) Scaling of the damping time with the wave period for the 1D MHD simulated case \citep[crosses; see][]{wan18} and that for the theoretical case (diamonds) same as in (a). The solid and dashed lines are the best fits to the theoretical and simulated data, respectively.
}
        \label{fig:tps}
\end{figure}

The approximately linear scaling between damping time and wave period has been revealed from empirical measurements of slow-mode waves in flare loops by multi-instrumental observations (see Sect.~3.2). Numerical simulations and linear theory based on 1D MHD models showed that this scaling relationship can be interpreted by the combined effect of thermal conduction and compressive viscosity \citep{ofm02,mend04,pan06,sig07}. The large scattering of data points (see Fig.~\ref{fig:wpf}b) may be due to the observed loops of different plasma parameters (e.g., in $n_0$ and $T_0$). Figure~\ref{fig:tps} compares the results obtained from the dispersion relations and nonlinear MHD simulations. Here we calculated the wave period and damping time for the combined thermal conduction and viscosity using $P=P_{\rm cond}+P_{\rm visc}-P_0\approx P_{\rm cond}$ and $\tau=1/(\omega_i^{\rm cond}+\omega_i^{\rm visc})$. Our tests based on Eq.~(\ref{eq:dsp}) showed that the additive property of linear dissipative processes works sufficiently well even in the regime of higher dissipation in hot loops with the physical parameters considered here and the infinite magnetic field approximation. The effects of nonlinearity and non-zero plasma-$\beta$ assessed, for example, in the thin flux tube approximation, on this estimation require further verification. The numerical model guided by SDO/AIA observations is same as that used in \citet{wan18}. A good agreement between the theoretical and simulated predictions is found (see Fig.~\ref{fig:tps}b). The curve for thermal conduction alone tends to be flattening at higher harmonics (see Fig.~\ref{fig:tps}a) indicating its damping saturated in the strong conduciton regime (i.e., $\tau\propto{P^0}$ for shorter periods). This implies that the viscosity is more efficient in dissipating the higher harmonics on small scale (or a fixed longitudinal mode in short loops), while the thermal conduction remains a dominant role in dissipating the fundamental mode on large scale. \citet{sig07} also showed a similar example (see the top-right panel of their Fig.~1). In addition, this characteristic of viscosity distinct from thermal conduction in the strong dissipation regime also accounts for its important role in suppressing the development of nonlinearity \citep{wan18}. Some nonlinear MHD simulations with no viscosity show velocity and density oscillations with tremendously large nonlinear effects \citep[see][]{mend04,sig07,fang15}, inconsistent with the AIA and SUMER observations. This also suggests that the inclusion of viscosity is essential in modeling slow-mode waves in hot flare loops. \\

\begin{figure}
        \centering
        \includegraphics[width=1.0\textwidth]{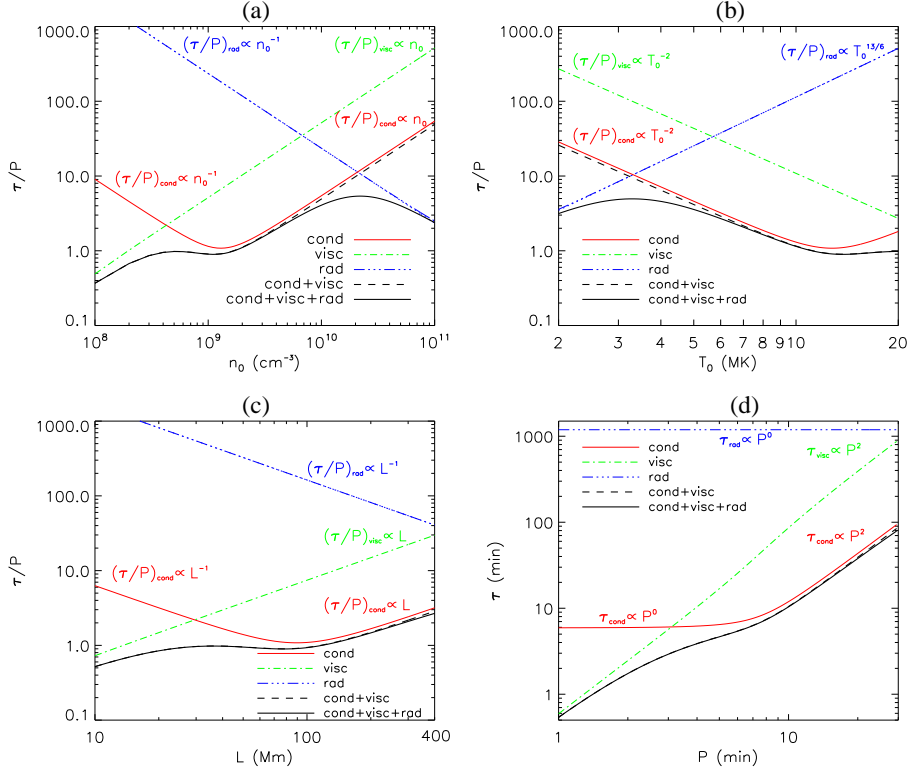}
        \caption{Variations in ratio of damping time to wave period with respect to (a) density ($n_0$), (b) temperature ($T_0$), and (c) loop length ($L$) in the cases with thermal conduction, compressive viscosity, optically thin radiation, and their combinations. The loop has the equilibrium $T_0$=9 MK, $n_0$=$2.6\times 10^9$ cm$^{-3}$, and $L$=180 Mm when one parameter varies. In (a)-(c), the dotted line represents the approximated solution for radiation (Eq.~\ref{equ:tpr}), which is nearly overlaid with the dot-dot-dashed line for the numerical solution. (d) Variations of damping time with wave period for the different dissipation mechanisms.
 }
        \label{fig:ntl}
\end{figure}

\noindent
(3) Optically thin radiation

Following \citet{dem04a}, a dimensionless parameter quantifying the effect of radiative loss on wave damping, namely the radiation ratio is defined as
\begin{equation}
r=\frac{P_0}{\tau_{\rm rad}}=\frac{(\gamma-1)\rho_0^2\chi T_0^{\alpha}P_0}{\gamma p_0}=\frac{(\gamma-1)n_0^2 \Lambda(T_0)P_0}{\gamma p_0},
\end{equation}
where $P_0=\lambda/C_s$, $\tau_{\rm rad}$ is the radiation timescale, $\Lambda(T)$ is the radiative loss function. According to the piecewise powerlaw approximation of \citep{rosn78}, $\Lambda(T)=10^{-17.73}T^{-2/3}$ erg~cm$^3$~s$^{-1}$ for $T\approx{2-10}$ MK. This approximation gives $\chi=4.6\times 10^{29}$, $\alpha=-2/3$, and $\tau_{\rm rad}=371 T_0^{5/3}/n_0$. For the fundamental mode of $\lambda=2L$, the radiation ratio $r$ can be expressed as,
\begin{equation}
 r=2.7\times 10^{-3}\left(\frac{n_0P_0}{T_0^{5/3}}\right)=3.55\times{10^{-7}}\left(\frac{n_0L}{T_0^{13/6}}\right).
 \label{equ:rad}
\end{equation}
To maintain the loop in thermal equilibrium, a constant heating function is assumed to balance the radiative cooling, i.e., $H_0=\rho_0^2\chi T_0^\alpha$, during the wave perturbations. The following dispersion relation can be derived from the linearized MHD equations \citep[see][]{dem04a,pan06,sig07},
\begin{equation}
 \omega^3 - i\left(\frac{r\alpha\gamma}{P_0}\right)\omega^2- (C_s^2 k^2)\omega- i(2-\alpha)\frac{rC_s^2 k^2}{P_0}=0.
 \label{equ:vdsp}
\end{equation}
In the case when the dependence of heating function on density and temperature (i.e., $H=H(\rho, T)$) and its perturbations due to slow-mode waves are considered, a misbalance between heating and cooling processes near the perturbed equilibrium may lead the wave dynamics to different regimes including growing, quasi-stationary, and rapidly damping (see Sect.~\ref{sss:whc}). Dispersion relation~(\ref{equ:vdsp}) on $\omega$ can be solved numerically for a fixed timescale $P_0$. In typical hot coronal loops (e.g., $T_0=6-10$ MK and $n_0={10^9-10^{10}}$ cm$^{-3}$), the thermal ratio is small ($r<$0.1; see Table~\ref{tab:der}). By transforming dispersion relation~(\ref{equ:vdsp}) into the form,
\begin{equation}
\omega^2 -C_s^2k^2 = i\frac{r}{\omega P_0}[\alpha\gamma\omega^2 + (2-\alpha)C_s^2k^2],
\end{equation}
and considering $\omega\approx{C_sk}$ and $\omega_r\gg\omega_i$ when $r\ll$1, it can reduce to
\begin{equation}
\omega \approx C_sk + i\frac{r}{2P_0}[\alpha(\gamma-1)+2]. 
  \label{equ:svdp}
\end{equation}
The simplified dispersion relation~(\ref{equ:svdp}) agrees with that derived by \citet{sig07}. We then have
\begin{eqnarray}
 P &=& \frac{2\pi}{\omega_r}\approx P_0, \\
 \tau &=& \frac{1}{\omega_i}\approx \frac{2P_0}{r[\alpha(\gamma-1)+2]} \propto{P_0^0}, \\
 \frac{\tau}{P} &\approx& \frac{2}{r[\alpha(\gamma-1)+2]} \propto \frac{T_0^{3/2-\alpha}}{n_0L}=\frac{T_0^{13/6}}{n_0L}.  \label{equ:tpr}
\end{eqnarray}

Note that because the presence of the parameters $d$, $\epsilon$, and $r$ in Eq.~(\ref{equ:cdsp}), (\ref{equ:vdpa}), or (\ref{equ:vdsp}) is in the form of $dP_0$, $\epsilon P_0$, and $r/P_0$ that are independent of $P_0$, the solutions of the corresponding dispersion relation for a certain harmonic (e.g., $k=\pi/L$ for the fundamental mode) are irrelevant to the choice of timescale $P_0$ (or lengthscale $L_s=P_0 C_s$ in some studies), although the values of $d$, $\epsilon$, and $r$ depend on $P_0$. Thus, one should be cautioned when comparing the results from different studies on wave dissipations, where the different timescales or lengthscales may be used.

We compare the individual and combined effects of the different dissipative terms on the wave damping based on the linear MHD theory. Table~\ref{tab:der} lists the values of thermal ratio, viscous ratio, and radiation ratio for the physical parameters in a wide range. Using these parameters we calculated the dependences of the ratio between damping time and wave period for the fundamental mode on the density, temperature, and loop length (Figs.~\ref{fig:ntl}a-c), where the characteristic power-law scalings for the individual mechanisms are marked. We find the following major features: (1) Thermal conduction damping is dominant over the other mechanisms for the typical hot coronal loops of $n_0\approx 10^9-10^{10}$ cm$^{-3}$, $T_0\approx 5-10$ MK, and $L\gtrsim 100$ Mm; (2) Damping by viscosity becomes comparable to or even more efficient than thermal conduction for the loops of lower density ($n_0<10^9$ cm$^{-3}$), higher temperature ($T_0>10$ MK), and shorter length ($L<100$ Mm) corresponding to the strong thermal conduction regime, while the effect of radiation is negligible in such a condition; (3) Radiative damping becomes comparable to or even more important than thermal conduction for the loops of higher density ($n_0>10^{10}$ cm$^{-3}$) and lower temperature ($T_0<5$ MK) corresponding to the weak thermal conduction regime, while in this case the effect of viscosity is negligible. Figure~\ref{fig:ntl}d shows the dependence of damping time on wave period for the fundamental mode in the loops of different sizes, showing the similar damping features to the case for different harmonics in a loop of the fixed length (see Fig.~\ref{fig:tps}a). Feature (1) supports the conclusion in \citet{ofm02}\footnote{Note that the density $n_0=1.5\times{10^9}$ cm$^{-3}$ instead of $5\times{10^8}$ cm$^{-3}$ was used in the simulations of \citet{ofm02}. The latter number was due to a typo.} that thermal conduction is the dominant damping mechanism for slow-mode waves in typical hot coronal loops. Feature (2) can account for the conclusion in some studies that the damping times by viscosity and thermal conduction alone are comparable \citep{mend04,sig07,abed12}. This is because the uncommon low densities with $n_0\lesssim 5\times{10^8}$ cm$^{-3}$ were used in all the cases of these studies, resulting in the thermal ratio $d\gg$0.1 \citep[e.g., $d\approx{0.3-10}$ in][]{sig07} -- a condition that has the thermal conduction damping less efficient. Feature (3) is in line with the favorable conditions for radiative damping (i.e., in the denser and/or cooler loops) found by \citet{pan06} and \citet{alg15}. 

In addition, we notice from Fig.~\ref{fig:ntl} that the ratio of damping time to wave period predicted by the combined dissipation mechanisms has a minimum about 1 for the typical hot loops, close to the averages for the SUMER and BCS observations \citep[see][]{wan11}. However, if the observed loops do not satisfy the physical condition that predicts the minimum $\tau/P$, other damping mechanisms could be invoked, such as the anomalous transport \citep{wan15,wan18}, or the wave-caused heating/cooling imbalance (see Sect.~\ref{sss:whc}). We provide an example here of anomalous transport conditions. \citet{wan07} measured seven hot loop oscillations with coordinated SUMER and Yohkoh/SXT observations, and obtained the average physical parameters $(\tau/P)_{\rm obs}=1.3\pm 0.7$, $T_0=6.6\pm 0.4$ MK, $n_0=(7.4\pm 3.3)\times 10^9$ cm$^{-3}$, and $L=116\pm 44$ Mm. Using  Eq.~(\ref{equ:dcgs}) we estimate the thermal ratio $d=0.022\pm 0.009$ and then derive the theoretical ratio $(\tau/P)_{\rm the}=3.5\pm 1.7$ from the curve for thermal conduction in Fig.~\ref{fig:tpd}b. Since the result of $d<{0.1}$ implies that the thermal conduction damping dominates over the viscous damping ($\tau_{\rm cond}/\tau_{\rm visc}\sim{0.1}$; see Eq.~\ref{equ:ctv}), the result of $(\tau/P)_{\rm the}\gtrsim 2(\tau/P)_{\rm obs}$ suggests that the dissipation by thermal conduciton is insufficient to account for the observed rapid damping. If we assume that the viscosity coefficient is anomalously enhanced by an order of magnitude compared to the classical value, the viscous damping time would become comparable to the conduciton damping time, thus the combined effect of the two mechanisms could explain the observations.

\subsection{Wave-induced heating/cooling imbalance}
\label{sss:whc}
In addition to a range of magnetically driven phenomena, the solar corona is a natural thermodynamically active medium. Indeed, the hot coronal plasma exists only due to a subtle balance between continuous loss of energy by optically thin radiation and some unknown yet heating mechanism counteracting it. Moreover, those plasma heating and cooling processes are likely to depend on the background plasma parameters differently \citep[see e.g.][]{dem15, klim15}, so that a destabilization of the initial quasi-steady state by some external compressive perturbation can lead to an effective energy exchange between the perturbation and the background plasma. In this section, we discuss damping of slow magnetoacoustic waves due to the wave-induced misbalance between plasma heating and cooling processes, focusing on derivation and estimation of the characteristic damping time and misbalance time scales for various physical conditions of the corona.

In the presence of some unspecified heating $H$ and optically thin radiative cooling $\mathcal{L}$, both determined by the plasma parameters such as density and temperature and thus both affected by the wave-caused perturbations of plasma, the energy equation can be written as
\begin{equation}\label{eq:energy_misbal}
C_{V}\frac{dT}{dt} - \frac{k_\mathrm{B}T}{m\rho}\frac{d\rho}{dt}=-Q+\frac{\kappa_\|}{\rho}\frac{\partial^2T}{\partial z^2},
\end{equation}
where $C_{V}=(\gamma-1)^{-1}k_\mathrm{B}/m$ is the specific heat capacity, $m$ is the mean particle mass, $\kappa_\|$ is the field-aligned thermal conductivity, and the combined heat/loss function $Q = \mathcal{L} - H$. We consider an isothermal plasma equilibrium, in which $Q_0=0$ and so the right-hand side of Eq.~(\ref{eq:energy_misbal}) is zero. Having the plasma perturbed by a compressive, in particular, slow-mode wave, different dependences of the functions $\mathcal{L}$ and $H$ upon plasma parameters can cause non-zero values of the perturbed heat/loss function $Q$, that is referred here to as a wave-induced heating/cooling misbalance. We note here that the radiative damping mechanism (3) considered in Sect.~\ref{sss:nme} represents a particular case of this more general heating/cooling misbalance process for a constant heating function $H=H_0$. In this case, the heating function has no effect on the wave dynamics, and its role reduces to maintaining the initial thermal equilibrium in the system.

\begin{SCfigure}
        \centering
        \includegraphics[width=0.65\textwidth]{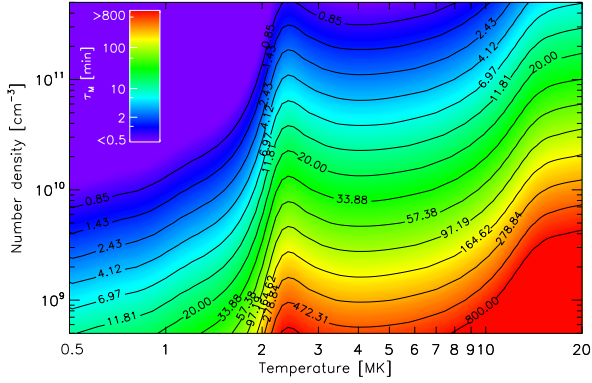}
        \caption{Thermal misbalance time $\tau_\mathrm{M}$ in the solar corona as a function of coronal plasma density and temperature, given in Eq.~(\ref{eq:dr_solim_misbal}) and estimated using CHIANTI atomic database v. 9.0.1 for optically thin radiative losses and a guessed coronal heating function ($H(\rho,T)\propto\rho T^{-3}$). The greenish color shows values of $\tau_\mathrm{M}$ approximately from 10 to 100 minutes. Adapted from \citet{kolot20}.
        }
        \label{fig:tau_misbal}
\end{SCfigure}

Applying the infinite magnetic field approximation, within which the set of MHD equations governing the slow-mode wave dynamics reduces to the one-dimensional hydrodynamic continuity equation, Euler equation, ideal gas state equation, and the energy equation, and linearizing it around the initial equilibrium, we obtain the following third-order differential equation for plasma density perturbed by a slow-mode wave in such a thermally active plasma
\begin{multline}\label{eq:wave_eq_misbal}
\frac{\partial^3 \rho}{\partial t^3} -  \gamma \frac{k_\mathrm{B} T_0}{m} \frac{\partial^3 \rho}{\partial t \partial z^2}
=\frac{\kappa_\|}{\rho_0 C_\mathrm{V}}\left(\frac{\partial^4 \rho}{\partial z^2\partial t^2} - \frac{k_\mathrm{B}  T_0 }{m} \frac{\partial^4 \rho}{\partial z^4} \right)\\
 -\frac{Q_{T}}{C_\mathrm{V} }\left(\frac{\partial^2 \rho}{\partial t^2} - \left[1-\frac{\rho_0}{T_0}\frac{{Q_{\rho}}}{Q_{T}}\right]\frac{k_\mathrm{B}  T_0 }{m}   \frac{\partial^2 \rho}{\partial z^2}\right),
\end{multline}
describing dynamics of two acoustic modes and one thermal mode, with $Q_{T} \equiv \left( \partial Q  / \partial T \right)_{\rho}$ and $Q_\mathrm{\rho}\equiv(\partial Q/\partial \rho)_T $ \citep[see][]{zav19}. As in the zero-$\beta$ plasma, slow magnetoacoustic waves do not perturb the magnetic field and thus it has no effect on the wave dynamics apart from determining the propagation direction and 1D nature of the wave \citep{duck20}, the dependence of the heating function on the magnetic field is omitted in Eq.~(\ref{eq:wave_eq_misbal}). Writing the density perturbation in Eq.~(\ref{eq:wave_eq_misbal}) as $\rho\propto e^{i(kz-\omega t)}$ and applying approximation of weak non-adiabaticity, i.e. assuming processes of thermal conduction and heating/cooling misbalance are slow in comparison with the wave period, \citet{kolot19} derived the dispersion relation for slow-mode waves in the plasma with heating/cooling misbalance,
\begin{equation}\label{eq:dr_misbal}
\omega^2 = C_\mathrm{s}^2 k^2\left\{1-i\omega^{-1}\left[\frac{\gamma - 1}{\gamma}\frac{1}{\tau_\mathrm{cond}}+\frac{\tau_1-\tau_2}{\tau_1\tau_2}\right]\right\},
\end{equation}
where $C_\mathrm{s}=\sqrt{\gamma k_\mathrm{B}T_0/m}$ is the sound speed and
\begin{align}
&\tau_{\mathrm{cond}}={\rho_0 C_\mathrm{V}k^{-2}}/{\kappa_\|},\nonumber\\
&\tau_{1}={\gamma C_\mathrm{V}}/\left[{Q_{T}-(\rho_0/T_0)Q_{\rho}}\right],\nonumber\\
&\tau_{2}={C_\mathrm{V}}/{Q_{T}},\nonumber
\end{align}
are the characteristic time scale of the parallel thermal conduction, and those describing rates of change of the heat/loss function $Q$ with plasma density and temperature. Considering real wavenumber $k$ and complex cyclic frequency $\omega = \omega_\mathrm{r}+i\omega_\mathrm{i}$ with $\omega_\mathrm{i}\ll\omega_\mathrm{r}$, Equation~(\ref{eq:dr_misbal}) can be resolved as
\begin{align}\label{eq:dr_solre_misbal}
&\omega_\mathrm{r} \approx C_\mathrm{s} k,\\
&\omega_\mathrm{i} \approx -\frac{1}{2}\left(\frac{\gamma - 1}{\gamma}\frac{1}{\tau_\mathrm{cond}}+\frac{1}{\tau_\mathrm{M}}\right),\label{eq:dr_solim_misbal}
\end{align}
where $\tau_\mathrm{M} = {\tau_1\tau_2}/{(\tau_1-\tau_2)}$ can be referred to as a characteristic time of the heating/cooling misbalance.  For $\tau_1>\tau_2$, $\tau_\mathrm{M}>0$ so that the discussed effect of the heating/cooling misbalance contributes into the wave damping.  Thus, in the considered limit of weak non-adiabaticity, the effects of the parallel thermal conduction and of the wave-caused heating/cooling misbalance on the slow-mode wave damping are additive. In other words, naturally present thermodynamical activity of the solar corona can lead to the enhanced damping of slow-mode waves in comparison with that caused by the thermal conduction alone \citep[see also][]{kums16,nak17}. The other case with $\tau_1<\tau_2$ corresponds to the regime of suppressed damping or even thermal over-stability due to an effective gain of the energy from the medium. In particular, on the linear stage such energy gain may lead to formation of quasi-periodic patterns \citep{zav19}, and to formation of the trains of self-sustained pulses on the nonlinear stage \citep{rob10,zav20}. 

\begin{figure}
        \centering
        \includegraphics[width=0.8\textwidth]{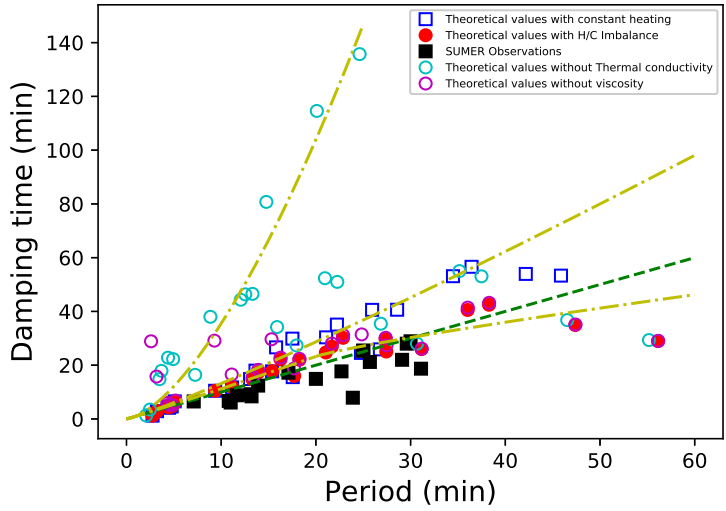}
        \caption{Variation of damping time ($\tau$) versus period ($P$): (i) filled-red circles (with thermal conduction$+$compressive viscosity$+$heating/cooling imbalance); (ii) blue rectangles (with thermal conduction$+$compressive viscosity$+$constant heating); (iii) cyan circles (only with compressive viscosity$+$heating/cooling imbalance); (iv) pink circles (only with thermal conduction$+$heating/cooling imbalance); (v) filled-black rectangles (observed SUMER oscillations). Three dot-dashed yellow lines show the fittings to the theoretical data points for cases (i)-(iii), respectively. The dark green-dashed line is $\tau$=$P$ line. Adapted from \citet{pras20}. }
        \label{fig:scthc}
\end{figure}

Typical values of the thermal misbalance time $\tau_\mathrm{M}$ leading to the slow-mode wave damping, estimated for various values of the coronal temperature and density, are illustrated in Fig.~\ref{fig:tau_misbal}. The temperature range shown in Fig.~\ref{fig:tau_misbal} covers emission formation temperatures of such observational instruments as SDO/AIA, SOHO/SUMER, Hinode/EIS, and Yohkoh/BCS. For this, we modeled the optically thin radiative loss function $\mathcal{L}(\rho,T)$ using CHIANTI atomic database v. 9.0.1 \citep{der97, der19}, and parametrized the unknown coronal heating function as $H(\rho,T)\propto \rho^a T^b$ \citep[see e.g.][]{dah88,iba93} with some guessed heating model $a=1$ and $b=-3$, ensuring the other (thermal) mode described by Eq.~(\ref{eq:wave_eq_misbal}) is stable. In contrast to $\tau_{\mathrm{cond}}$ which grows with density and decreases with temperature, for the chosen heating model $\tau_{\mathrm{M}}$ decreases with density and grows with temperature, that could indicate the domination of different physical mechanisms in the slow-mode wave damping in different plasma conditions and require further dedicated investigation. In particular, one of the important implications of the discussed slow-mode wave damping mechanism is that it can occur even in isothermal waves. In that regime, the wave is not subject to damping by thermal conduction at all, while the heating and cooling processes and, hence, the wave-caused misbalance between them can affect the wave amplitude via the perturbations of plasma density. Note that the isothermal waves can also be damped by compressive viscosity and leakage.

Recently, \citet{pras20} analyzed the slow-mode wave damping for the range of loop lengths $L=50-500$ Mm, temperatures $T_0=5-30$ MK, and densities $n_0=5\times10^9-5\times10^{11}$ cm$^{-3}$, based on a new dispersion relation derived from linearized MHD equations including thermal conductivity, compressive viscosity, radiation, and unknown heating term along with the consideration of heating/cooling imbalance. Figure~\ref{fig:scthc} shows the damping time vs. wave period for various kinds of theoretically estimations. They found that the predicted scaling law can match better to the observed SUMER oscillations for an assumed heating function $H(\rho, T)\propto\rho^{-1/2}T^{-3}$ when the heating/cooling imbalance is taken into account.

\section{Applications of coronal seismology with slow-mode waves}
\label{sct:acs}
\subsection{Transport coefficients}
 \label{sst:tc}
The dissipation of slow-mode waves is closely related to transport processes in the coronal plasma (see Sect.~\ref{sss:nme}). \citet{nak00} interpreted propagating, quasi-periodic disturbances observed in warm coronal loops as the propagating slow-mode waves based on a theoretical model that includes the nonlinearity and various dissipation effects, and suggested to use the waves as a diagnostic tool for MHD coronal seismology. For example, the dissipative coefficient (related to compressive viscosity and thermal conduction) can be estimated by comparing the measurement of wave amplitudes as a function of the distance along the loop with those predicted by evolutionary equation. They found that the classical viscosity coefficient needs to be enhanced by one to two orders of magnitude to account for the observed decay when considering the viscosity alone. This is understandable because for those coronal loops with typical parameters $T_0$=1.6 MK and $n_0=5\times 10^8$ cm$^{-3}$, as well as the wave period $P=300-900$ s, the thermal ratio $d$ is estimated to be in the range 0.02$-$0.07 corresponding to the weak thermal conduction regime, where the damping rate due to thermal conduction is higher than that due to viscosity by an order of magnitude (see Eq.~\ref{equ:ctv}). This implies that the viscosity enhancement is required in order to have the effect of dissipation comparable to that of thermal conduction. \citet{dem03} obtained the similar results based on a linear wave theory.

\citet{vand11} first observationally determined the effective adiabatic index $\gamma_{\rm eff}$ (or the polytropic index) in coronal loops from the relative amplitudes of density and temperature perturbations, and estimated the thermal conduction coefficient from their phase lag. \citet{krish18} statistically measured the propagating slow-mode waves in sunspot fan loops from 30 different active regions observed with SDO/AIA and obtained $\gamma_{\rm eff}=1.05-1.58$ with a mean of $1.1\pm 0.1$, consistent with that measured by \citet{vand11}. The fact of $\gamma_{\rm eff}\sim{1}$ implies that the waves propagate at the nearly isothermal sound speed (in the strong thermal conduction regime with the expected thermal ratio $d\gg{0.1}$). Compared to the actual thermal ratio $d_0=0.014$ estimated using Eq.~(\ref{equ:dcgs}) from the parameters of observed fan loops (with $n_0\approx 2\times 10^9$ cm$^{-3}$, $T_0\approx 1$ MK, and $P\approx 180$ s), this suggests that for these conditions the thermal conduction coefficient needs to be significantly enhanced compared to the classical value. We here estimate the effective thermal ratio $d_e$ from $\gamma_{\rm eff}$ based on the polytropic approximation of $p=K\rho^{\alpha}$, where $K$ is a constant and $\alpha$ is the polytropic index, which gives the estimate of the phase speed as \citep[see][]{wan18},
\begin{equation}
 V_p\approx\left ( \frac{\partial{p} }{\partial \rho } \right )^{1/2}=\left(\frac{\alpha p_0}{\rho_0}\right)^{1/2}=\left(\frac{\alpha}{\gamma}\right)^{1/2}C_s.
\label{eq:cp} 
\end{equation}
Using this equation for $\alpha=1.1$ we derive $V_p\approx 0.81 C_s$, and then by using the relation between $V_p$ and $d$ shown in Fig.~\ref{fig:tpd}a, we obtain $d_e=0.18$. The result of $d_e/d_0=13$ implies that to account for the measured polytropic index the thermal conductivity needs to be enhanced by an order of magnitude.

\begin{figure}
        \centering
        \includegraphics[width=1.0\textwidth]{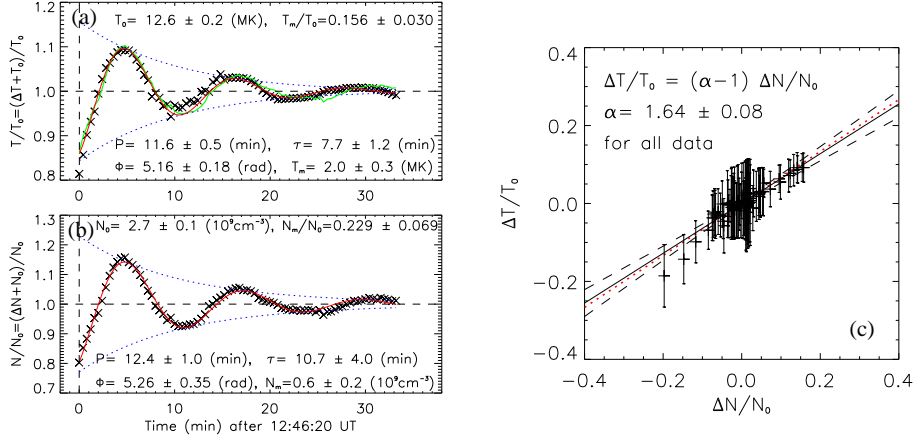}
        \caption{Variations of (a) temperature and (b) electron density, normalized to the corresponding trend for a region at the leg of a hot flaring loop observed with SDO/AIA on 2013 December 28 in AR 11936. The red solid line indicates the best fit to a damped sine function. The green solid curve in (a) is the predicted temperature from the observed relative density based on the adiabatic assumption. (c) Measurement of the polytropic index from the scatter plot of relative density against relative temperature. The solid line is the best fit to the data. The dotted line stands for $\gamma$=5/3. Panels (a) and (b) are from \citet{wan18} and Panel(c) from \citet{wan15}. }
        \label{fig:tcsp}
\end{figure}

In the following we exemplify some applications of the coronal seismology technique in determination of transport coefficients based on SDO/AIA observations of slow-mode oscillations in flaring coronal loops in combination with the linear wave theory and nonlinear MHD simulations. It is well known that dissipation of the slow magnetoacoustic waves by thermal conduction leads to a phase shift ($\Delta{\phi}$) between temperature and density perturbations \citep[e.g.][]{owe09}. Considering the case that thermal conduction is the only dissipation source the following relations for $\Delta{\phi}$ can be derived from the linearized energy equation \citep[see][]{wan18},
\begin{eqnarray}
{\rm tan}\,\Delta{\phi}&=&\frac{2\pi\gamma{d}\left(\frac{C_s}{V_p}\right)^2/(1+\chi^2)}{1-2\pi\gamma{d}\left(\frac{C_s}{V_p}\right)^2\chi/(1+\chi^2)},  \label{eq:phg} \\
(\gamma-1) {\rm cos}\,\Delta{\phi}&=& \frac{A_T}{A_n}\left[1-2\pi\gamma{d}\left(\frac{C_s}{V_p}\right)^2\chi/(1+\chi^2)\right], \label{eq:amg}
\end{eqnarray}
where $\gamma=5/3$ and the frequency $\omega=\omega_r + i\omega_i$ can be calculated from the dispersion relation (\ref{equ:cdsp}) for a fixed wavenumber $k$ for a standing mode, the phase speed $V_p=\omega_r/k$,
$\chi=\omega_i/\omega_r$, $A_T=T_{1m}/T_0$ is the relative amplitude of perturbed temperature, and $A_n=n_{1m}/n_0$ the relative amplitude of perturbed density. Under the assumption of weak dissipation approximation ($V_p=\omega_r/k\approx{C_s}$ and $\chi\approx{0}$) the above equations reduce to \citep[see][]{vand11,wan15,krish18},
\begin{eqnarray}
  {\rm tan}\,\Delta{\phi}&=& 2\pi\gamma d,      \label{eq:pha} \\
  (\gamma-1) {\rm cos}\,\Delta{\phi}&=& \frac{A_T}{A_n} ~~[\approx\alpha-1]. \label{eq:ama}
\end{eqnarray}
Numerical analyses show that the estimate of $\Delta{\phi}$ from Eq.~(\ref{eq:pha}) has a relative difference of $\leq{7}\%$ with respect to the solution of Eq.~(\ref{eq:phg}) without the above approximations for the weak damped oscillations with $\tau/P\geq{2}$ (corresponding to $d\leq{0.04}$ that gives $\Delta{\phi}\leq{24}^\circ$).

Note that the relation ${A_T}/{A_n}=\alpha-1$ is strictly valid only under the polytropic assumption. In this assumption, the ratio between $A_T$ and $A_n$ can be practically linked to the polytropic index $\alpha$ by
\begin{equation}
\frac{T_1}{T_0}=(\alpha-1)\frac{n_1}{n_0}, \label{eq:fgm}
\end{equation}
or
\begin{equation}
\alpha(t)=\frac{A_T(t)}{A_n(t)}+1,  	   \label{eq:tgm}
\end{equation}
where $T_1$ and $n_1$ are the perturbed temperature and density, $A_T(t)$ and $A_n(t)$ are their instantaneous relative amplitudes normalized to the corresponding trends $T_0(t)$ and $n_0(t)$. Under the polytropic approximation, the polytropic index $\alpha$ can be measured by fitting the scaling between $T_1/T_0$ and $n_1/n_0$ (after first removing their phase shift $\Delta{\phi}$). This technique has been applied to a number of observed and simulated data sets \citep{vand11,wan15,wan18,wan19,krish18}. The time-dependent polytropic index can be obtained using Eq.~(\ref{eq:tgm}) by determining $A_T(t)$ and $A_n(t)$ with the Hilbert transform \citep[e.g.][]{real19}. The phase shift $\Delta{\phi}$ between the temperature and density oscillations can be measured using the cross correlation from their relative time profiles of $T_1/T_0$ and $n_1/n_0$ \citep[e.g.][]{wan18}. In addition, the instantaneous phase shift $\Delta{\phi(t)}$ can be determined using the Hilbert transform.

\begin{figure}
        \centering
        \includegraphics[width=1.0\textwidth]{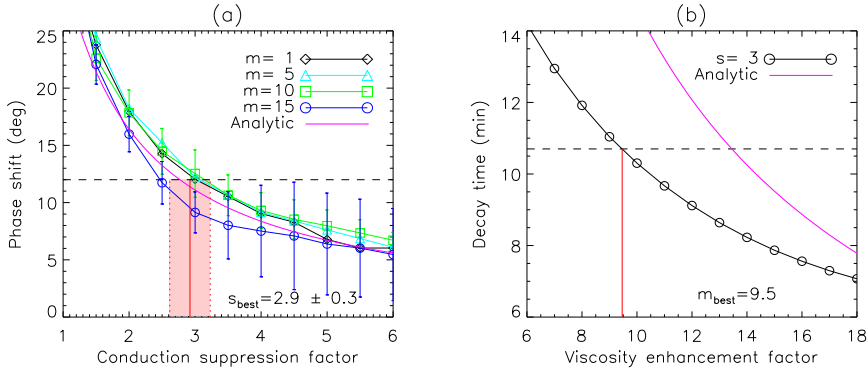}
        \caption{Determination of transport coefficients using parametric simulations based on a 1D nonlinear MHD model \citep[from][]{wan19}. (a) Dependence of the phase shift between the density and temperature oscillations on the conduction suppression factor ($s$). The vertical red line indicates the determined $s$-factor from the observed phase shift. (b) Dependence of the decay time on the viscosity enhancement factor ($m$). The vertical red line indicates the determined viscosity enhancement factor from the observed phase shift. The pink solid curves in both plots are the solution of linear theory. }
        \label{fig:sch}
\end{figure}

\citet{wan15} analyzed a longitudinal oscillation event observed with SDO/AIA, and found convincing evidence for the strong suppression of thermal conduction in hot flaring loops. Figure~\ref{fig:tcsp} shows that the observed temperature and density variations are nearly in-phase and the measured polytropic index is close to $\gamma=5/3$. This result agrees with the prediction by Eq.~(\ref{eq:ama}), i.e., $\Delta{\phi}=0$ infers $\alpha=\gamma$, and vice visa. For this event the estimated thermal ratio $d\approx 0.07$ (see the note of Table~\ref{tab:der}), suggests that the damping by thermal conduction is supposed to dominate over that by compressive viscosity. The finding of the thermal conduction suppression, thus, implies a significant enhancement of viscosity invoked to explain the rapid decay of the observed waves. Considering the wave dissipation by viscosity alone, the effective viscosity coefficient can be obtained in the form of the observables by eliminating $P_0$ and $\epsilon$ from Eqs.~(\ref{equ:vis})$-$(\ref{equ:vtt}) as, 
\begin{equation}
 \eta_0^{cs}=\frac{3\gamma p_0\tau}{8\pi^2(\tau/P)^2+2}
  = \frac{3\gamma k_Bn_0T_0\tau}{4\pi^2(\tau/P)^2+1}, \label{eqvis}
\end{equation}
where $P$ and $\tau$ are the observed wave period and decay time, respectively, and $p_0=2n_0k_BT_0$. Using the measured thermal and wave parameters, \citet{wan15} found $\eta_0^{cs}/\eta_0=15$, where $\eta_0$ is the classical Braginskii viscosity coefficient. 

Based on 1D nonlinear parametric simulations including thermal conduction and compressive viscosity, \citet{wan19} refined on the method of determining the transport coefficients using a two-step procedure: (1) determine the effective thermal conduction coefficient from the observed phase shift between temperature and density perturbations because this physical parameter is insensitive to the unknown viscosity (see Fig.~\ref{fig:sch}a), as it was demonstrated by \citet{wan19}; (2) with the loop model of the thermal conduction coefficient obtained in step 1, determine the effective viscosity coefficient from the observed decay time using the parametric modeling (see Fig.~\ref{fig:sch}b). \citet{wan19} applied this new coronal seismology technique to the wave event studied in \citet{wan15}, and obtained improved results that the classical thermal conduction coefficient is suppressed by a factor of about 3 and the classical viscosity coefficient is enhanced by a factor of 10 in the hot flaring loop. By applying coronal seismology to the observations of coexisting kink and slow-mode oscillations of coronal loops, \citet{nis17} obtained the plasma-$\beta$ to be about 0.1$-$0.3 and the effective adiabatic index $\gamma_{\rm eff}\approx{5/3}$. Their result of $\gamma_{\rm eff}$ could also suggest a significant suppression of thermal conduction in the analyzed hot loop hosting the slow-mode waves.  

The observationally-determined transport coefficients have significant implications for our understanding of the thermodynamic processes in hot flaring loops. The thermal conduction suppression suggests that the flaring loop should cool much slower than expected from the classical Spitzer conductive cooling. This mechanism may provide an alternative interpretation for long-duration events (LDEs) observed in SXR and EUV radiations \citep[e.g.][]{taka00,qiu12}. The thermal conduction suppression also suggests a weaker chromospheric evaporation \citep{karp87}, and so may lead to a phenomenon that hot coronal loops tend to be underdense compared to the hydrostatic prediction \citep{wine03, real14}. In addition, numerical simulations by \citet{wan18} revealed that the viscosity enhancement plays an important role in efficiently dissipating higher harmonic components of an impulsively-generated disturbance, self-consistently explaining the quick formation of the observed fundamental standing mode. Some mechanisms for the anomalies of transport coefficients have been suggested. For example, the suppression of thermal conduction may be caused by nonlocal conduction \citep[e.g.][]{karp87}, or turbulent scattering \citep[e.g.][]{jiang06,bian18}. The enhancement of compressive viscosity may be attributed to turbulence such as Bohm diffusion and eddy viscosity \citep{bohm49, holl88}. However, a mechanism that can simultaneously account for the both effects is still unknown.

\subsection{Heating function}
\label{sst:hf}

The presence of oscillations has long been found in the models of the evolution of flaring loops \citep[e.g.][]{Jakimiec1992a,warren2002,Bradshaw2013a}, although initially not specifically addressed. Although the observational evidence of oscillations is still debated, especially for stellar flares \citep{vand2016}, specific numerical modeling promptly connected impulsive heating to low-frequency slow-mode waves leading to quasi-periodic perturbations \citep{nak04,tsik04,selw05}. Some observational studies suggested that these oscillations are possibly triggered by small flares near one loop footpoint \citep{wan05}, or indirectly through the modulation of non-thermal electron beams \citep{Takahashi2017a,nak18}. Early loop modeling efforts were devoted to investigate pulsations detected with SOHO/SUMER \citep{tar05, tar07}. More complex modeling has been used to investigate the mechanism for QPPs in multi-component magnetic systems and events, such as flares with CMEs \citep{Takahashi2017a}. The propagation and reflection of slow-mode waves were produced also with 2D MHD models of a flare with a heat pulse located at the loop footpoint \citep{selw07,fang15}, and with 3D MHD models where a pulsed velocity driver was used to excite oscillations associated with upflows caused by a storm of small heat pulses \citep{selw09,wan13,prov18}.

The origin of quasi-periodic pulsations detected in flare light curves is still unclear \citep[see][for a review]{vand2016}. In the hypotheses that QPPs are caused by slow magnetoacoustic modes in low-$\beta$ limit, i.e., acoustic modes, confined in a coronal loop, and that the loops evolve symmetrically with respect to the apex, detailed hydrodynamic modeling shows that large amplitude oscillations (20\% in density) can be triggered in flare light curves if the duration of the heat pulse ($\Delta t_H$), when considering a symmetric heating along the loop, is shorter than the loop sound crossing time ($\tau_s$) at the flare maximum:
\begin{equation}
\Delta t_H < \tau_s = \frac{2L_h}{C_0}  \sim 5 \frac{L_{Mm}}{\sqrt{0.1 ~  T_{MK}}} ~~~~(\rm s)
\label{eq:tau_s}
\end{equation}
where $C_0$ is the isothermal sound speed, $L_h$ is the loop half-length ($L_{Mm}$ in units of Mm), and $T_{MK}$ the maximum loop temperature in units of MK \citep{real16,real19}. During the initial phase of the flare, the temperature increases steeply while the coronal density is still quite low. Even considering saturation effects \citep[e.g.][]{Cowie1977a}, the thermal conductivity is very effective and the conduction time along the loop is shorter than the sound crossing time. The temperature of the medium where the wave propagates can be considered uniform and the temperature disturbance is expected to be very small, at least initially \citep{real19}. Therefore, the estimate of $\tau_s$ under the assumption of isothermal sound speed appears to be reasonable.

\begin{SCfigure}  
 \centering
 \includegraphics[width=0.6\textwidth]{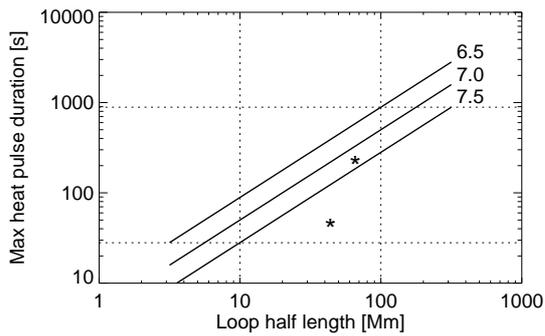}
\caption{Maximum heat pulse duration to trigger quasi-periodic pulsations vs the loop half length (solid lines) for the 3 labelled loop maximum temperatures ($\log T$). The central dashed box bounds maximum durations for typical solar coronal loop lengths (10 to 100 Mm). The values obtained from detailed modeling of pulsations observed in a SDO/AIA observation of a transient loop system are marked with stars \citep[see][]{real19}. }
\label{fig:pulse_dur}
\end{SCfigure}

The physical reason for this condition is a combination of dynamic and energetic effects. The heat pulse drives explosive evaporation of plasma from the chromosphere and a strong supersonic pressure front travels upwards along the loop from both footpoints (for the uniform heating case), and then bounces back. If the heat pulse stops before the return pressure front arrives back at the footpoints, the sudden deficit of heating creates a pressure dip there, which acts as an elastic restoring force and sustains the front sloshing \citep{real16}.

In Fig.~\ref{fig:pulse_dur} the possible values of $\Delta t_H$ are in the region below the lines that track $\tau_s$ from Eq.~(\ref{eq:tau_s}) for three temperatures. In general, we expect heat pulses to bring the plasma temperature to values $\log T[K] > 6.5$, likely even higher than 10 MK. For solar flaring loops of typical half-lengths $L_h=10-100$ Mm, heat pulses will trigger pulsations if their duration is below $\sim 30$~s for small loops, $\sim 1000$~s for long loops. It is therefore more likely to detect pulsations in long flaring loops. As an example of direct evidence for this scenario, large-amplitude pulsations have been detected, and modeled in detail, in a hot transient loop system observed with SDO/AIA, and in particular in loops with estimated half-length of 40 Mm and 60 Mm. The heat pulse duration that best fits data with a loop hydrodynamic simulation is 30~s and 150~s, respectively \citep{real19}.

\begin{figure}    
 \centering
   \includegraphics[width=0.40\textwidth,height=0.6\textheight]{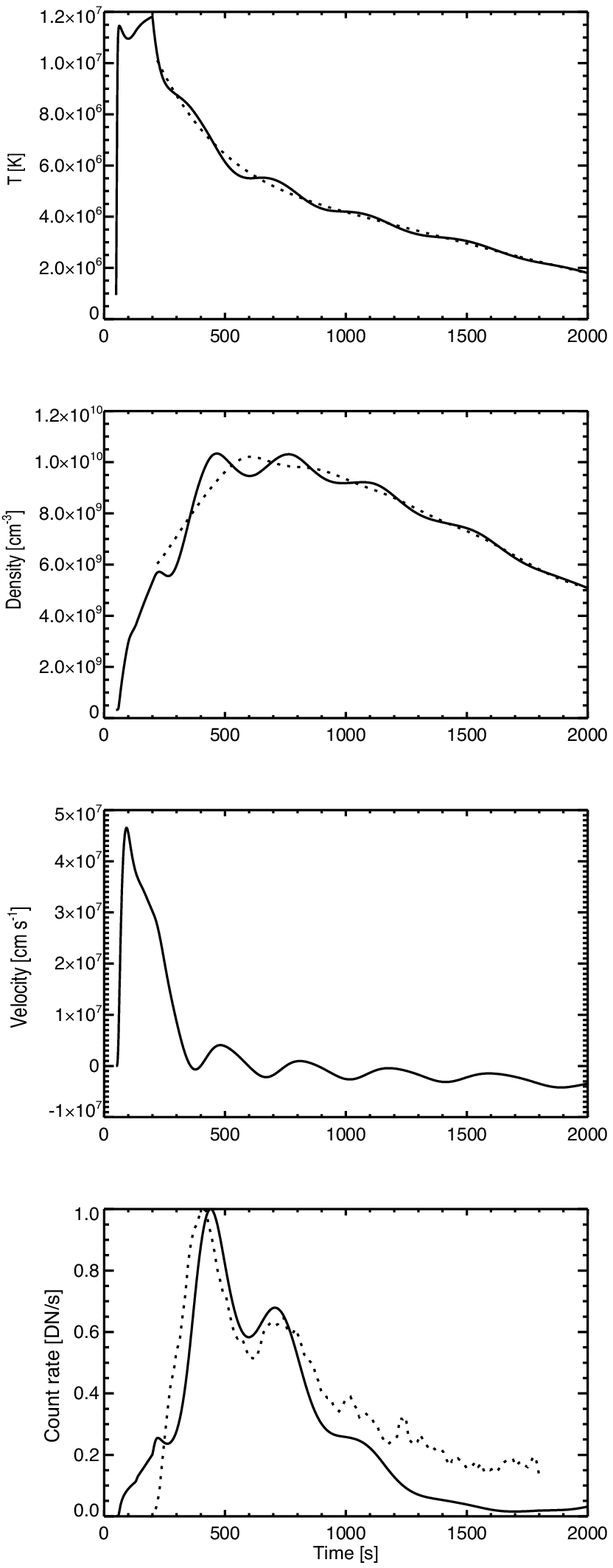}
   \includegraphics[width=0.40\textwidth,height=0.6\textheight]{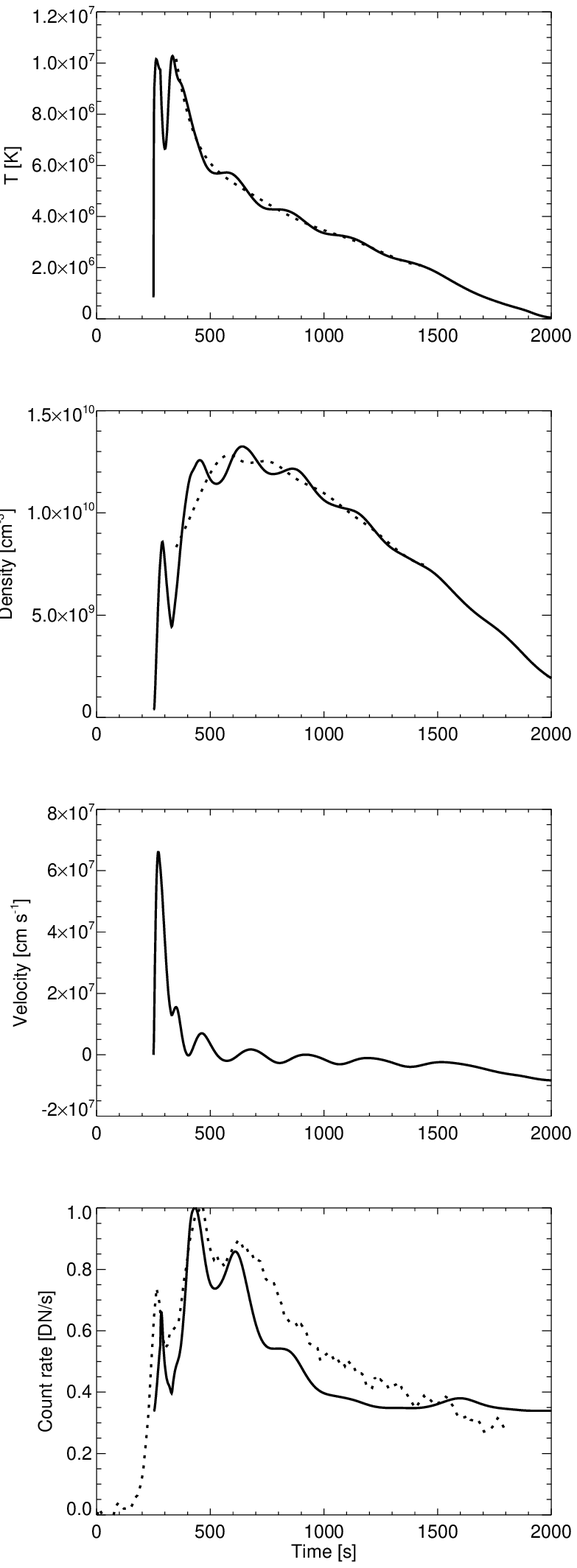}
\caption{Results of hydrodynamic modeling of (left) a coronal loop heated with a pulse deposited uniformly in the loop and (right) a shorter loop heated with a pulse deposited symmetrically at the both footpoints: from top to bottom, evolution of the average temperature, density, velocity, and normalized light curve (solid lines) synthesized in the AIA 94~\AA\ channel in low loop segment of 30~Mm (left) and 20~Mm (right). The smoothed trends of the temperature and density variations are also shown (dashed lines). The observed light curves (dashed lines) are also shown for comparison. Adapted from \citet{real19} }
\label{fig:mod_aia}
\end{figure}

Equation~(\ref{eq:tau_s}) can also be applied to estimate the period of the pulsation, but using the time-averaged temperature. In fact, the pulsation train is detected while the loop is very bright, i.e., very dense, well after the heat pulse is over and the plasma is already cooling. Figure~\ref{fig:mod_aia} shows the evolution obtained from hydrodynamic loop modeling that provides good fits to two pulsed light curves observed in the loop system mentioned above. The pulsation periods $\sim 6$ min for the longer loop (left) and $\sim 3$ min for the other one (right) broadly corresponds to a temperature of $\sim 7-10$~MK. According to Fig.~\ref{fig:mod_aia}, the general shape of the light curves can be reproduced by a single loop model. This suggests that in highly transient coronal events the loops evolve mostly as a whole with no effective structure into fine strands, and the heat pulse may involve most of the loop cross-section.

The detailed hydrodynamic modeling of \citet{real19} also demonstrated a higher level of diagnostics of the heating function. The light curve on the right of Fig.~\ref{fig:mod_aia} shows an initial spike that is not observed in the other. The hydrodynamic loop models can reproduce the both cases with and without the spike. The basic difference between the two models is the location where the heat pulse is deposited: in the case with the spike the heat pulse is deposited near the loop footpoints, in the other case it is deposited higher in the coronal part of the loop. A low pulse location may indicate the presence of non-thermal electron beams hitting the chromosphere, while a higher one may indicate a different mechanism of heat deposition, perhaps direct magnetic reconnection. The details of the pulsations then could become an important diagnostic tool to constrain impulsive coronal heating.

This level of diagnostics is possible on the Sun, because we can spatially resolve the coronal structures and detect pulsations at precise locations. However, similar QPPs are detected also from spatially unresolved observations of stellar flares. It has been shown that the same kind of hydrodynamic modeling of flaring loops is able to reproduce, even quantitatively, the pulsations that modulate SXR flare light curves in protostars \citep{real18}. These flares are typically very long-lasting (1 day or more) and the pulsation periods are proportionally long (hours). Since the data do not allow for strict constraints on the heating function, the assumption of a heat pulse released in the stellar corona is good enough. The model coherently describes a long flare with pulsations observed in a star of the Orion cluster, with a very long pulse-heated loop. The observed modulations, especially when with a large amplitude, can be reproduced by integrating over the loop length if assuming a small part of the loop is invisible (e.g., near one footpoint). Some other possibilities might be explored to explain the wave signals detected in the total flux, such as the asymmetric loop geometry or the projection effect.

The detailed solar and stellar loop modeling that led to wave triggering conditions (Eq.~\ref{eq:tau_s}) and heating diagnostics in \citet{real16} and \citet{real19} has focused on symmetric heating and loop evolution. We do not expect large differences if the pulse location is not symmetric as in \citet{real19} model but still high in the solar or stellar corona, because heat will be conducted rapidly along the loop getting to both footpoints with a slight delay, and making the evolution anyway rather symmetric. The case of a strongly asymmetric heat location, i.e., concentrated at only one of the loop footpoints, has also been investigated \citep{selw05,selw07,tar05,tar07,fang15}. Such asymmetric pulse will produce the fundamental standing mode, instead of the second harmonic \citep{tar05,tar07,wan18}, and the condition (\ref{eq:tau_s}) for triggering pulsations should still hold, by replacing the loop half length $L_h$ with the total loop length. However, using a similar hydrodynamic model with impulsive heating, \citet{tsik04} found the occurrence of second-harmonic standing slow-mode waves near the peak and decay phases of a flare, independent of the location of heat deposition in the loop. Their results could be attributed to the choice of a very different heating condition, i.e., $\Delta{t}_H\gg\tau_s$, under which the loop plasma is gradually heated to a super-hot state of the peak temperature up to 30 MK. The second harmonic waves may be excited by strong upflows at the both footpoints due to chromospheric evaporation.

\subsection{Diagnostic techniques based on multi-harmonics}
\label{sst:mh}
Remarkable developments took place in the last decade with observations of multiple harmonics of slow magnetoacoustic oscillations in the solar and stellar loops \citep[see][and references therein]{wan09,sri10,kumm11,sri13,krish14,kum15,pugh15}. These observations have been used to infer the information on the longitudinal structuring and transport processes by mean of MHD seismology \citep[e.g.][]{sri13,krish14}.

\subsubsection{Effect of longitudinal structuring on the period ratio}
In a homogeneous loop model, the ratio between the periods of first two harmonics ($P_{1}/2P_{2}$) of slow (or acoustic) modes must be equal to one. However, in the more realistic longitudinally structured system, it is shifted to the lower values. This feature of the waves can be utilized as a seismological tool to diagnose the nature of the local corona. For an isothermal loop of the half-length $L_h$ where the propagation speed $C_s$ of slow (or acoustic) waves and acoustic cut-off frequency ($\Omega_c$, depending upon the gravitational stratification) remain constant, the period ratio can be estimated as \citep{Mac06},
\begin{equation}
{\frac{P_{1}}{2P_{2}}} = {\Bigg(  {\frac{1+\frac{\Omega_c^{2}L_h^{2}}{\pi^{2}C_s^{2}}}{1+\frac{4\Omega_c^{2}L_h^{2}}{\pi^{2}C_s^{2}}}} \Bigg)}^{1/2} = {\Bigg(  {\frac{1+\frac{1}{4\pi^{2}}{(\frac{L_h}{\Lambda_{c}})^{2}}}{1+\frac{1}{\pi^{2}}{(\frac{L_h}{\Lambda_{c}})^{2}}}} \Bigg)}^{1/2}, \label{equ:prt}
\end{equation}
where the acoustic cut-off frequency is related to the pressure scale height ($\Lambda_{c}$) as $\Omega_c=C_{s}/2\Lambda_{c}$. It is obvious that $0.5\leq P_{1}/2P_{2}\leq{1}$.  The density stratification causes the period ratio $P_{1}/2P_{2}$ to fall off from unity significantly in the longer loops ($L_h>\Lambda_{c}$). For the shorter and isothermal loops, if we ignore the gravity then $\Omega_c$=0, and the period ratio $P_{1}/2P_{2}$ will be equal to one. In the non-isothermal loops (considering temperature increases from the loop base to apex) where $C_s$ and $\Omega_c$ vary along the loop, the departure of $P_{1}/2P_{2}$ from unity will happen even when the effect of the gravitational stratification is not taken into account.

Applying Eq.~(\ref{equ:prt}) to an observation of slow-mode waves with $P_{1}/2P_{2}$=0.92 and $L_h$=33 Mm, \citet{sri10} estimated the density scale height $\Lambda_c\sim21$ Mm in a coronal loop observed in EUV. This result may suggest that the observed loop departs from the hydrostatic equilibrium condition and/or the loop is non-isothermal because the isothermal model predicts $\Lambda_c\sim70$ Mm at the measured coronal temperature $T=1.6$ MK, which is much larger than that derived by coronal seismology.

Another application case we review here is associated with stellar observations. \citet{sri13} found the first evidence for multiple harmonic slow-mode waves in the post-flaring loops of the corona of Proxima Centauri using XMM-Newton observations. They detected the periodic pulsations of 1261 s and 687 s during the flare decay phase in the SXR band (0.3$-$10 keV). The flare loop was estimated to have a length of 75 Mm with the peak temperature of 33 MK and the decay-phase temperature of $\sim$7 MK on average. By interpreting the observed two periodicities in terms of first two harmonics of the standing slow modes, \citet{sri13} inferred from the period ratio $ P_1/2P_2=0.91$ the density scale height $\Lambda_c=23$ Mm in this stellar loop system. The derived density scale height is much smaller than that ($\sim$300 Mm) expected for the 7 MK plasma, suggesting that the observed period ratio may result from other effects, e.g. temperature and magnetic stratifications.

\citet{Luna12} developed a model considering the effect of magnetic stratification with the uniform density in a semi-circular loop geometry, and derived the period ratio for slow modes as
\begin{equation}
{\frac{P_{1}}{2P_{2}}} = 1-{\frac{15\beta_{f}}{6+5\beta_{f}}}(\Gamma-1),  \label{equ:prm}
\end{equation}
where $\Gamma$ is the loop-expansion factor ($\Gamma=1$ for a non-expanding loop), and $\beta_{f}$ is the plasma-$\beta$ at the footpoint of the loop. Equation~(\ref{equ:prm}) clearly indicates that magnetic stratification (i.e., the loop expanding with height) has the same effect on the period ratio as gravitational density stratification, in contrast with the behavior for kink modes where the shifts of the period ratio due to the density and magnetic stratifications counteract each other \citep{Andries09}. It also shows that in a very low-$\beta$ environment the effect of magnetic stratification on the period ratio is less important than that of density stratification \citep{abed11,Luna12}. However, some observational studies showed that the departure of observed $P_{1}/2P_{2}$ from unity is much larger than the model-predicted when considering density stratification alone in a coronal loop system \citep{sri10,kumm11}. This suggests that the observations need to be explained by some other effects such as temperature gradient along the loop \citep{abed11,abed12} and non-ideal MHD effects (see the next section).

\subsubsection{Effect of wave dissipation on the period ratio}

Some theoretical studies have also assessed the influence of non-adiabatic damping such as compressive viscosity, thermal conduction, and optically thin radiation on the period ratio of the first two harmonics \citep[e.g.][references cited there]{mac10,kumn11,abed12}. This approach not only helps to improve the diagnostic of longitudinal structuring in a corona loop using the period ratio, but also raises its potential of obtaining the information on non-ideal conditions.

\citet{mac10} first analytically examined the effects of thermal conduction and compressive viscosity on the period ratio of slow-mode waves based on a 1D uniform loop model. From Fourier analysis ($e^{i(\omega{t}-kz)}$) of the linearized MHD equations, they derived the dispersion relation as
\begin{equation}
\Omega^3 -i(\mathcal{V}+\gamma\mathcal{D})\Omega^2 - (1+\gamma\mathcal{V}\mathcal{D})\Omega + i\mathcal{D}=0, \label{eq:dsp}
\end{equation}
where $\Omega=\omega/kC_s$, $\mathcal{D}=dP_0C_sk$, $\mathcal{V}=(4/3)\epsilon{P_0}C_sk$, $P_0=2L/C_s$, and $L$ is the loop length. Here to be convenient for discussion, the definitions of thermal ratio $d$ and viscous ratio $\epsilon$ are same as in Sect.~\ref{sss:nme}. By solving the dispersion relation (\ref{eq:dsp}) numerically with $k=\pi/L$ and $k=2\pi/L$, the dimensionless frequencies for the fundamental mode ($\Omega_1$) and the second harmonic mode ($\Omega_2$) can be obtained, respectively, giving the period ratio as $P_1/2P_2=Re(\Omega_2)/Re(\Omega_1)$.

\begin{figure}
         \centering
         \includegraphics[width=0.9\textwidth]{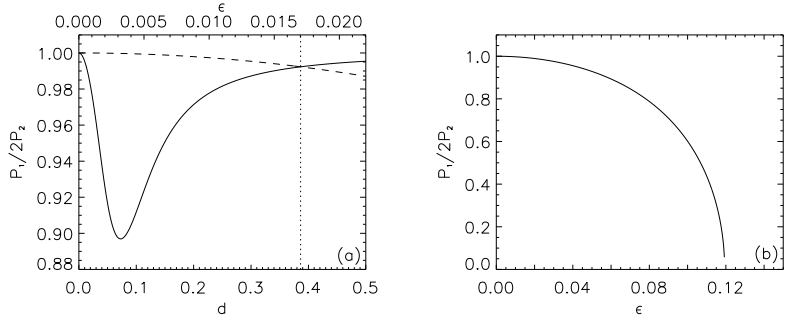}
         \caption{(a) Period ratio $P_1/2P_2$ as a function of the thermal ratio $d$ (solid line). The dashed line represents $P_1/2P_2$ against the viscous ratio ($\epsilon$). Note that $d/\epsilon=22.7$ is a constant. The vertical dotted line indicates the crossing point (at $d=0.386$) between the two curves. (b) Period ratio $P_1/2P_2$ as a function of the viscous ratio $\epsilon$. Note that the plots shown here are similar to those in \citet{mac10} but different in definitions of $d$ and $\epsilon$. }
         \label{fig:pr}
 \end{figure}

Figure~\ref{fig:pr}a shows the behavior of the period ratio with thermal ratio $d$ (solid line) in the absence of compressive viscosity. The period ratio has a minimum of $P_1/2P_2$=0.897 at $d_m=0.0726$, and returns to unity when $d$ decreases to 0 or becomes sufficiently large. Based on the loop model with physical parameters varying in a wide range, \citet{mac10} concluded that for both warm (1$-$2 MK) EUV loops and hot ($6-10$ MK) SUMER loops, the effect of thermal conduction on the period ratio is negligible. We argue that their conclusion is supported in the former case but appears not to be supported by observations in the latter case. For example, for the event studied by \citet{sri10}, a coronal loop of length $L$=66 Mm, temperature $T$=1.6 MK, and density $n=10^9$ cm$^{-3}$ gives $d$=0.015, leading to $P_1/2P_2$=0.99. However, for typical SUMER hot loops with $L$=100$-$200 Mm, $T$=6$-$10 K, and $n=10^9-10^{10}$ cm$^{-3}$, we find $d$=0.019$-$0.14 in a range that well covers $d_m$ where the departure of $P_1/2P_2$ from unity reaches the maximum. We notice that the reason leading to the conclusion of \citet{mac10} is that they assumed a fixed loop pressure for all conditions. This would result in the density for the hotter loops to be significantly underestimated (e.g., $n=2\times{10}^8$ cm$^{-3}$ at $T$=10 MK for the given $p_0$=0.55 dynes cm$^{-2}$), so leading to $D=2\pi{d}\gg{1}$ for the fundamental mode. This condition implies that $P_1/2P_2\rightarrow{1}$ (see Fig.~\ref{fig:pr}a).

In the absence of thermal conduction (i.e., $\mathcal{D}=0$), the dispersion relation (\ref{eq:dsp}) can be solved with $\mathcal{V}_1=8\pi\epsilon/3$ and $\mathcal{V}_2=16\pi\epsilon/3$  to give the period ratio
\begin{equation}
\frac{P_1}{2P_2}=\left(\frac{1-(64/9)\pi^2\epsilon^2}{1-(16/9)\pi^2\epsilon^2}\right)^{1/2}, \label{eq:prv}
\end{equation}
where the viscous ratio needs to satisfy $\epsilon<3/(8\pi)$ to keep $P_2$ as a real number. Figure~\ref{fig:pr}b shows that $P_1/2P_2$ decreases monotonically with $\epsilon$. As the ratio $d/\epsilon\approx$23 is a constant (see Sect.~\ref{sss:nme}), we compare the dependence of $P_1/2P_2$ on $d$ and $\epsilon$ in the same plot (Fig.~\ref{fig:pr}a). It indicates that the effect of viscosity on the period ratio is negligible ($P_1/2P_2>0.993$) when thermal conduction damping dominates over viscous damping (when $d<0.37$; see Fig.~\ref{fig:tpd}b). This includes the cases of typical SUMER oscillations. For the same reason mentioned above (i.e., a significant underestimate of density for the hotter loops), \citet{mac10} concluded that the viscous ratio is large enough for very hot and short SUMER loops to produce a significant effect on the period ratio. However, evidently this is not the case in the observed events. For example, a hot loop with $T$=10 MK, $L$=50 Mm, and $n=2\times10^9$ cm$^{-3}$ gives $\epsilon$=0.0165 which leads to a period ratio $P_1/2P_2$=0.993. It is only for super-hot (e.g., $T>20$ MK) small postflare loops (but with density not very high, e.g. $n<10^{10}$ cm$^{-3}$) that the effect of compressive viscosity on the period ratio may become important. For example, for a postflare loop of $T$=20 MK, $L$=20 Mm, and $n=5\times10^9$ cm$^{-3}$ it gives $\epsilon$=0.066, leading to $P_1/2P_2$=0.867.

In addition, the effect of radiation on the period ratio was also examined by some authors \citep{kumn11, abed12}. They found that the radiative damping is not the main dissipative agent that affects the wave modes and their period ratio. Compared to thermal conduction and compressive viscosity, the radiation has a negligible effect on the period ratio in both warm and hot coronal loops with or without consideration of temperature inhomogeneity.

\section{Conclusions and open questions}
\label{sct:coq}

We have presented a review of recent advances in observation and theory on studies of slow magnetoacoustic waves in AR coronal loops, focusing on the so-called ``SUMER oscillations" that are characterized by their associations with impulsive heating and signatures of long periods, large amplitudes, and quick decaying. New observations from SDO/AIA and Hinode/XRT discovered both standing and reflected propagating longitudinal intensity oscillations in flaring loops manifesting many features (e.g., wave periods, decay times, and triggers) in agreement with the SUMER oscillations (Sect.~\ref{sct:swhlp}). Numerical simulations based on the 1D MHD model (Sect.~\ref{sst:msw}) and 2.5D MHD model (Sect.~\ref{sst:mrw}) constrained by AIA observations suggested that a reflected propagating slow-mode wave tends to be produced when the thermal conduction damping is dominant in hot loops of a normal condition (i.e., with the classical transport coefficients), whereas a standing slow-mode wave tends to quickly form in an abnormal condition when compressive viscosity is enhanced and dominates in damping (e.g., due to anomalous transport). In the regime of strong thermal conduction, dissipation of higher harmonics becomes inefficient favoring the growth of nonlinearity and the existence of the higher harmonics that is necessary to sustain the reflected slow-mode waves in the form of pulse shape as observed. On the other hand, in the condition with the significantly enhanced viscosity (and possible suppression of thermal conduction), dissipation of the higher harmonics becomes efficient and the effects of nonlinearity are highly suppressed, favoring the formation of a standing wave. In addition, isothermal 3D MHD simulations revealed that realistic AR structures may play an important role in the quick formation of the standing slow-mode wave by impulsive events, possibly due to wave leakage related to transverse structuring in the curved geometry (Sect.~\ref{sst:msw}). This suggests that 3D MHD models with realistic initial and boundary conditions including non-ideal dissipation effects may be required to further improve our insights into the nature of observed slow-mode waves. 

The rapid decay is the most revealing aspect in the properties of observed slow-mode waves and has been intensively studied in theory since the discovery of SUMER oscillations. We have given an overview of various damping mechanisms based on nearly all the relevant published studies in the recent literature and summarized their analysis methods in Table~\ref{tab:dmp}. Overall, it can be concluded from these studies that the non-adiabatic dissipations by thermal conduction, compressive viscosity, and optically thin radiation are the major damping mechanisms for slow-mode waves (Sect.~\ref{sss:nme}). The dominant effect among these three mechanisms strongly depends on the loop physical condition (i.e., values of $n_0$, $T_0$, and $L$). Linear theories and MHD modeling suggest that thermal conduction dominates the damping of the fundamental mode in typical hot loops and the combined effect with compressive viscosity can account for the roughly linear scaling between damping time and wave period as revealed from solar and stellar observations. In addition, recent theoretical studies have shown that when considering the presence of certain unspecific heating mechanism, dependent on the plasma parameters (e.g., $n_0$ and $T_0$), the slow-mode waves will affect both the heating and radiative cooling, leading to the perturbed heating/cooling imbalance that may significantly change the behavior of the wave evolution (Sect.~\ref{sss:whc}). The possible role of heating/cooling imbalance in damping of the slow-mode wave in coronal loops of different physical conditions needs further exploration.

Seismological applications using slow-mode waves to probe the transport coefficients and heating function in hot flaring loops have been successfully developed as evident from the reviewed studies. The resultant finding of anomalous transport processes such as thermal conduction suppression and viscosity enhancement may provide new insights on the long-standing puzzle of the long-duration flaring events (Sect.~\ref{sst:tc}). The HD modeling has been used to diagnose impulsive heating in coronal loops by reproducing the QPPs detected in solar and stellar flares (Sect.~\ref{sst:hf}). It can provide constraints on the heating deposition including the location, duration, and heating rate. We expect that the state of the art 3D MHD modeling of realistic AR magnetic configuration will further improve the diagnostic power of forward modeling extending previous models.

The following are important questions and issues that have yet to be fully elucidated: 
\begin{enumerate}
\item What are differences in the physical condition that produces the standing and reflected propagating slow-mode waves found in observations of flaring loops?

\item How does the condition of coronal loops (e.g., when a loop is pre-existing and hot prior to the flare, or it is heated and forming during the flare) affect the nature of excited slow-mode waves?

\item \citet{nak19} proposed that the competition of nonlinear and dissipative effects could play an important role in generation of reflected propagating slow-mode waves. This idea needs to be tested with more samples in a proper method (e.g., considering the influence of the projection effect on Fourier decomposition analysis) and with numerical modeling.  

\item How to clearly differentiate a standing or reflected propagating slow-mode wave in observations? If an observed wave is transitioning from the propagating mode into the standing mode, how could one describe this feature quantitatively? Determination of the formation time of a standing wave from observations may allow the development of new applications in coronal seismology.
\end{enumerate}

SDO/AIA also discovered standing slow-mode waves in a non-flaring coronal fan loop system (Sect.~\ref{sct:swflp}), which was generated by the global EUV waves originating from a remote AR. The related open questions are
\begin{enumerate}
\item How is a standing slow mode excited in the coronal fan loop system, characterized by obviously asymmetric distributions in density and magnetic fields along the loop?

\item Why did the intensity oscillations, observed with no evident decay, last only about one period, then suddenly become undetectable?

\item The oscillations are detected with weak damping only at a few locations along the fan loop. Does it imply the presence of higher harmonics which are expected to damp faster than the fundamental mode?
\end{enumerate}

The kind of decaying long-period QPPs in solar and stellar flares have been connected with standing slow-mode waves (Sect.~\ref{sct:qpp}). This provides us with the possibility to explore the impulsive heating processes in the corona by the solar-stellar analogy. Some related open questions are 
\begin{enumerate}
\item Whether are these decaying harmonic type of QPPs in solar and stellar flares, when interpreted as slow-mode waves, mainly produced by direct modulations of the loop plasma or by indirect modulations of magnetic reconnection?

\item How to reliably determine whether the QPPs are related to the fundamental mode or to the second (or higher) harmonic? Are they generated by asymmetric heating (e.g., a flare at one footpoint) or symmetric heating (e.g., precipitation of energetic particles at both footpoints)?
\end{enumerate}

We envision that a statistical study of longitudinal oscillations of flaring loops based on AIA observations in combination with realistic multidimensional MHD modeling will be imperative to answer the above open questions regarding SUMER oscillations. The new EUV spectrometer SPICE onboard the Solar Orbiter mission \citep{mull20} including two 10 MK flare lines (Fe\,{\sc xviii} $\lambda$974.8 and Fe\,{\sc xx} $\lambda$721.5), will provide important Doppler velocity diagnosis of longitudinal waves complementary to imaging observations from SDO/AIA. In the near future, the Multi-Slit Solar Explorer (MUSE) will obtain EUV spectral images with a raster scan cadence 100 times higher than the present, critical for diagnosing the impulsive heating processes and related wave activity in coronal loops \citep{dep20}.

 \begin{acknowledgements}
This review is based upon the discussions with the members of the Science Team on ``Oscillatory Processes in Solar and Stellar Coronae" at the workshop at ISSI-BJ. The work of T.W. and L.O. was supported by NASA grants 80NSSC18K1131 and the NASA Cooperative Agreement NNG11PL10A to CUA. The work of T.W. was also supported by the NASA grant 80NSSC18K0668. L.O. also acknowledges support by the NASA grants NNX16AF78G. D.Y. is supported by the National Natural Science Foundation of China (NSFC, 11803005, 11911530690)and Shenzhen Technology Project (JCYJ20180306172239618). F.R. acknowledges support from Italian Ministero dell'Universit\`{a} e della Ricerca and contract ASI-INAF 2017-14-H.0. D.Y.K. acknowledges support from the STFC consolidated grant ST/T000252/1 and the budgetary funding of Basic Research program No. II.16. A.K.S. acknowledges his UKIERI Research for his scientific research. 
 \end{acknowledgements}

%
 

\bibliographystyle{spbasic}      
 \bibliography{Wang}


\end{document}